\documentclass[aps,pre,onecolumn]{revtex4}
\usepackage{graphicx, bm, bbm,amsmath, amssymb, epsfig}

\pdfoutput=1

\newcommand{\nn}{\noindent}

\newcommand{\bq}{\begin{align}}
\newcommand{\eq}{\end{align}}

\begin{document}
\title{Improved convection cooling in steady channel flows}
\author{Silas Alben}
\affiliation{Department of Mathematics, University of Michigan,
Ann Arbor, MI 48109, USA}
\email{alben@umich.edu}

\date{\today}

\begin{abstract}
We find steady channel flows that are locally optimal for transferring heat
from fixed-temperature walls, under the constraint of a fixed
rate of viscous dissipation (enstrophy = $Pe^2$), also the power needed to pump the
fluid through the channel. We generate the optima with net flux as a continuation parameter, starting from parabolic (Poiseuille) flow, the unique optimum at
maximum net flux. Decreasing the flux, we eventually reach optimal flows that
concentrate the enstrophy in boundary layers of thickness $\sim Pe^{-2/5}$ 
at the channel walls, and have a uniform flow with speed  $\sim Pe^{4/5}$ outside
the boundary layers. We explain the scalings using physical arguments with
a unidirectional flow approximation, and mathematical arguments using a
decoupled approximation. We also show that with channels of aspect ratio 
(length/height) $L$, the boundary layer thickness scales as $L^{3/5}$ and the outer flow speed scales as $L^{-1/5}$ in the unidirectional approximation. At the 
Reynolds numbers near the turbulent transition for 2D Poiseuille flow in air, we
find a 60\% increase in heat transferred over that of Poiseuille flow.  
\end{abstract}


\maketitle
\section{Introduction}

Heat transfer by forced convection is ubiquitous in residential, commercial, and
industrial settings, for example in
the heating and cooling of buildings, the cooling of power equipment and vehicles, 
and the manufacturing and processing of food, chemicals, metals, and other materials \cite{kreith2012principles}. 
The development of improved heat transfer technologies are 
an important part of
efforts to improve overall energy efficiency \cite{reay2013process}. A recent motivation is 
the rapid growth in 
cloud-computing data centers, recently estimated to account for 
about 2\% of the world's energy consumption \cite{dai2014optimum}. A significant fraction 
of this energy (5-50\%, depending on the data center)
is used not to run the computing equipment but to keep
it adequately cooled \cite{joshi2012energy,dai2014optimum}. The highest performing computing systems
are particularly dependent on efficient cooling by forced convection \cite{nakayama1986thermal,zerby2002final,mcglen2004integrated,ahlers2011aircraft,patterson2016energy,wagner2016test}. 

Heat transfer enhancement is the process of increasing the
heat transferred for a given amount of energy consumed to drive
the convecting flow. For example, the heated surface can be roughened 
to increase turbulence near the surface \cite{hart1985heat}, or vortex
generators can be added near the surface \cite{gee1980forced,tsia1999measurements,promvonge2010enhanced}. A related idea,
explored in a set of recent studies, is to place a rigid bluff body or an 
actively or passively flapping plate
near the heated surface \cite{fiebig1991heat,sharma2004heat,accikalin2007characterization,gerty2008fluidic,hidalgo2010heat,shoele2014computational,jha2015small,alben2015flag,wang2015dynamics}. Vorticity shed at the body
edges can enhance fluid mixing and thermal boundary layer disruption,
sometimes with relatively low energy cost.

Here we study the fluid flows that are optimal for heat transfer in
a 2D channel, a basic geometry. We ask: for a given amount of energy consumption, 
what fluid flow maximizes the rate of heat transfer from a heated surface? 
To simplify the problem, we focus on the convective cooling of the heated walls of a straight
planar channel, one of the most well-studied
geometries \cite{rohsenow1998handbook,shah2014laminar}, with early work in the 19th and early
20th centuries \cite{graetz1882ueber,leveque1928laws} and recent
applications such as channel shape 
optimization \cite{de2014shape}, and small-scale cooling where slip at the boundaries 
can play a role \cite{haase2015graetz}. The rate of heat transfer has
been calculated for developed or developing, laminar or turbulent flows.
Wall temperature held constant is the most common boundary condition, but
other conditions such as nonuniform wall temperatures or 
prescribed wall heat fluxes have been used \cite{rohsenow1998handbook,shah2014laminar}.

As a beginning optimization calculation, we focus on steady 2D
flows. It is conceptually  
straightforward to extend the approach to unsteady and/or 3D flows,
but the computational cost is considerably higher.
Recent work suggests that steady flows might obtain optimal heat
transfer in reduced models of 2D Boussinesq flows \cite{souza2015maximal,souza2015transport,souza2016optimal}.
Our main objective is to characterize the optimal steady flow structures as a
starting point towards understanding the optimal flows that can be 
generated by fixed or oscillating obstacles that generate vorticity or 
turbulence. With an understanding of the optimal steady flow structures, 
we will also obtain the corresponding scalings of heat transfer with the
(fixed) energy budget parameter. Recently, related work has studied the optimal 
flow for the mixing of a passive scalar in a fluid \cite{chien1986laminar,caulfield2001maximal,tang2009prediction,thomases2011stokesian,foures2014optimal,camassa2016optimal}, sometimes in a channel geometry.
Other work has studied optimal flow solutions for Rayleigh-B{\'e}nard convection 
\cite{waleffe2015heat,sondak2015optimal}, the transition to 
turbulence \cite{kerswell2014optimization,kaminski2014transient,pausch2015direct}
and heat transport from one solid boundary to another
\cite{hassanzadeh2014wall,souza2015transport,goluskin2016bounds,tobasco2016optimal,alben_2017}.
Alternative ways to improve heat transfer are to change the spatial and 
temporal configurations of heat sources and sinks
\cite{campbell1997optimal,da2004optimal,gopinath2005integrated}. 
Other optimal flows for heat transfer have been calculated with
alternative definitions or assumptions for the underlying fluid flow
\cite{karniadakis1988minimum,mohammadi2001applied,zimparov2006thermodynamic,chen2013entransy}.

\section{Problem setup \label{sec:Model}}

\begin{figure}
  \centerline{\includegraphics[width=15cm]
  {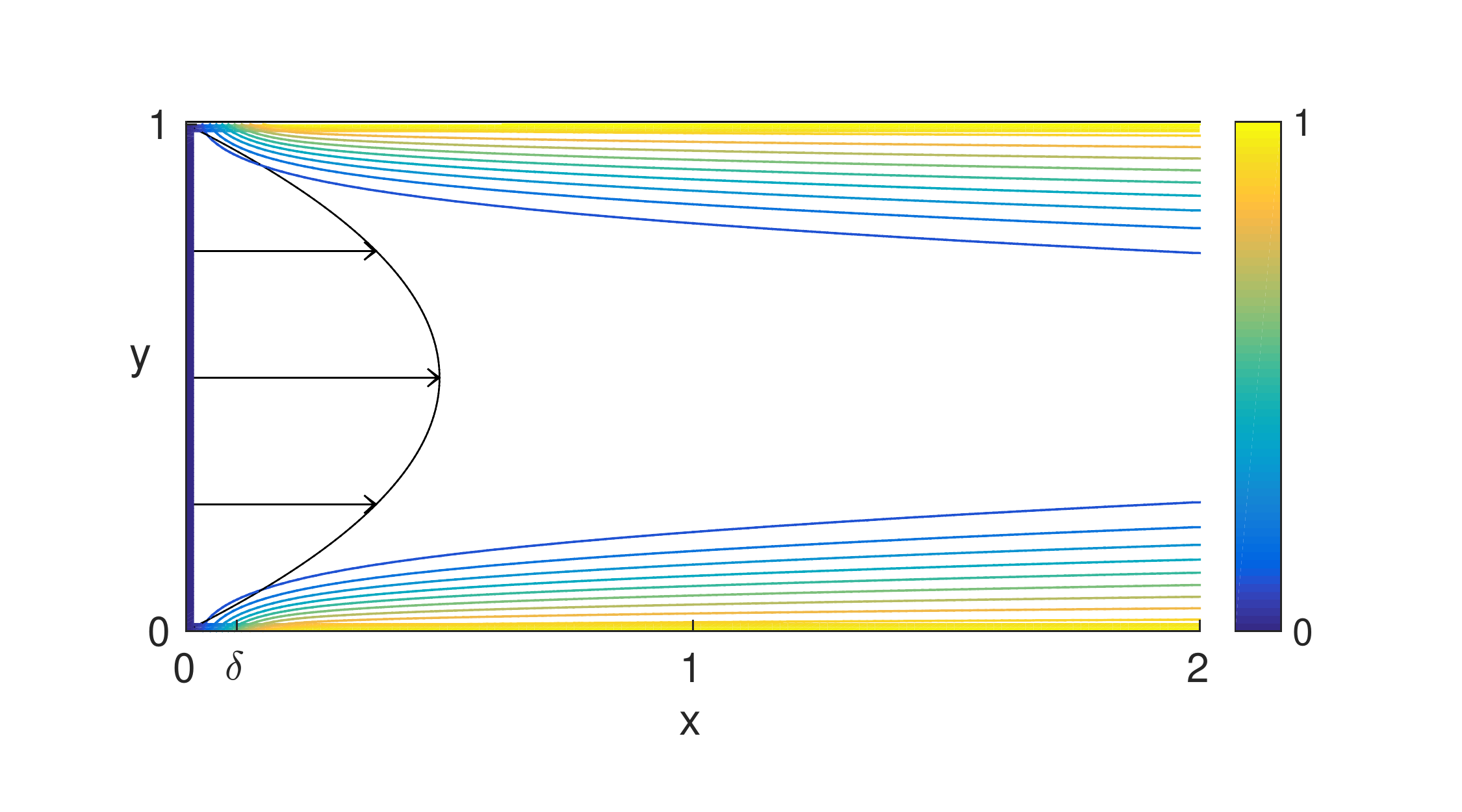}}
  \caption{Temperature contours for parabolic (Poiseuille) flow through a 
channel with height 1 and length 2. 
At the upstream boundary the fluid temperature is zero. On the top and bottom
walls the fluid temperature is $1 - e^{-x^2/\delta^2}$ with the
smoothing parameter $\delta$ set to 0.1. The flow is purely horizontal
with velocity profile $u(y) = 512\sqrt{6}(y - y^2)$, corresponding to 
dimensionless parameter values $Pe = 1024$ and $L = 2$.}
\label{fig:PoiseuilleFlowTempFig}
\end{figure}

We consider steady incompressible 2D fluid flows $\mathbf{u} = (u(x,y), v(x,y))$ in a channel of height $H$ and length $L_0$
occupying
$0 \leq x \leq L_0$, $0 \leq y \leq H$ (see Figure \ref{fig:PoiseuilleFlowTempFig}). The fluid temperature field $T(x,y)$ solves the steady advection-diffusion equation
\begin{align}
\mathbf{u} \cdot \nabla T - \kappa \Delta T &= 0, \label{AdvDiff}
\end{align}
\nn with $\kappa$ the thermal diffusivity of the fluid, equal to $k/\rho c_p$, where $k$ is the thermal conductivity, $\rho$
the fluid density and $c_p$ the specific heat capacity at constant pressure. At the inflow boundary, $x = 0$, cold fluid enters with temperature
$T = 0$. At the outflow boundary, $x = L_0$, we use an outflow boundary condition $\partial_x T = 0$, so the fluid temperature
maintains its value from slightly upstream of the boundary. On the top and bottom walls, $y = 0$ and $H$, the temperature is 
set to $T = 1-e^{\displaystyle -x^2/\delta^2}$,
where $\delta$ is a small smoothing parameter. The temperature is nearly
unity over most of the walls. We use $\delta$ to avoid a temperature singularity where the walls meet the upstream boundary, 
and ease the numerical resolution requirements. We work in the regime 
$\delta \ll  H, \delta \ll L_0$, and find little dependence of the solutions on 
$\delta$ in this limit.

The entering cold fluid heats up as it flows past the walls, so the exiting fluid
has a higher temperature. Integrating (\ref{AdvDiff}) over the channel and using the divergence theorem, we have
\begin{align}
\int T\mathbf{u} \cdot \mathbf{n} ds = \int \kappa \partial_n T ds. \label{AdvDiffBdy}
\end{align}
\nn where the integrals are over the entire boundary (top, bottom, upstream, and downstream). We assume no-slip conditions on
the top and bottom walls ($\mathbf{u} = 0$), leaving only the upstream and downstream sides to contribute to the left side of (\ref{AdvDiffBdy}).
This term gives the net advection of heat out of the channel (divided by $\rho c_p$). The right side gives the net heat flux into the channel
by conduction from the boundaries (divided by $\rho c_p$). The downstream boundary does not contribute due to the outflow condition 
$\partial_x T = 0$. For
flows with typical speed $\gg \kappa/H_0$ (the usual case in applications), conduction through the upstream boundary is slight and
most of the heat flux is from the top and bottom walls. 

\section{Viscous Energy Dissipation Constraint \label{sec:Enstrophy}}

The problem we consider is to find an incompressible fluid flow $\mathbf{u}$
that maximizes the (steady) rate of heat flux out of the hot walls,
\begin{align}
\int_0^{L_0} k\, \left(\partial_y T |_{y = H} - \partial_y T |_{y = 0} \right) dx, \label{Cond}
\end{align}
for a given rate of viscous dissipation per unit width in the out-of-plane
direction,
\begin{align}
\dot{E} = 2\mu W \iint e_{ij} e_{ij} dA = \mu W \iint (\Delta \psi)^2 dA. \label{disp}
\end{align} 
Here $\mu$ is viscosity, $W$ is the out-of-plane width (along which the flow and temperature are uniform), and $e_{ij} = (\nabla \mathbf{u} + \nabla \mathbf{u}^T)/2$ is the
symmetric part of the rate-of-strain tensor. The last term in (\ref{disp}) involves
the stream function $\psi$, which is defined for an incompressible 2D flow by 
$\mathbf{u} = -\nabla^\perp \psi = (\partial_y \psi, -\partial_x \psi)$. The second equality in
(\ref{disp}) is derived in \cite{lamb1932hydrodynamics} (article 329) and holds for
certain boundary conditions including those we will use here 
(we will have $v \equiv 0$ on the boundaries, which is sufficient for (\ref{disp})).

A (sufficiently smooth) incompressible flow solves the Navier-Stokes equations
\begin{align}
\rho\frac{D\mathbf{u}}{Dt} = -\nabla p + \mu \Delta \mathbf{u} + \mathbf{f} 
\quad ; \quad \nabla \cdot \mathbf{u} = 0, \label{NS}
\end{align}
\nn for a suitable forcing term $\mathbf{f}$, which represents a force per unit volume
applied over the fluid domain. If 
$\mathbf{f}$ takes the form of a $\delta$-distribution along a rigid or deforming no-slip 
surface, it represents the surface stress such boundaries apply to the flow, as in
the immersed boundary formulation \cite{Griffith07,shoele2014computational}. Much previous work
has studied the enhancement of heat transfer by vortices shed from 
rigid or flexible bodies in the flow \cite{gee1980forced,tsia1999measurements,promvonge2010enhanced,shoele2014computational,rips2017efficient}. A smooth 
$\mathbf{f}$ represents a distribution of forces spread over the flow
domain, and can approximate those applied within localized regions such
as the no-slip surfaces of vortex-generating obstacles, or powered fans. 

Taking the dot product on both sides of the first equation in (\ref{NS}) 
with $\mathbf{u}$ and integrating over the channel, we obtain the energy 
balance equation after some manipulations \cite{durst2008fluid}:

\begin{align}
\frac{d}{dt}\iint \frac{\rho|\mathbf{u}|^2}{2} dA + 
\int_0^H  \frac{\rho|\mathbf{u}|^2 }{2} u \Big{|}^{x = L_0}_{x = 0} dy +
\dot{E}/W
= \int_0^H p u\Big{|}^{x = L_0}_{x = 0} dy + \iint \mathbf{f} \cdot \mathbf{u} dA.
\label{EBal}
\end{align}
\nn We will assume steady flow, which is the same at the inflow and outflow
boundaries, so (\ref{EBal}) becomes
\begin{align}
\dot{E}/W = \int_0^H p u\Big{|}^{x = L_0}_{x = 0} dy + \iint \mathbf{f} \cdot \mathbf{u} dA.
\label{EBal1}
\end{align}
\nn The rate of work done by the volume distribution of forces $\mathbf{f}$ and
the pressure at the upstream and downstream boundaries equals the rate of
viscous dissipation (per unit width). Previous works \cite{gee1980forced,tsia1999measurements,promvonge2010enhanced,shoele2014computational,rips2017efficient} 
considered flows with the cost defined as the ``pumping power,'' the
first term on the right side of (\ref{EBal1}). Our definition of cost
is the rate of viscous dissipation, which by (\ref{EBal1})
is equal to both terms on the right side of (\ref{EBal1}): 
pressure at the boundaries and volume forces in the interior. Thus our cost is
the same as the ``pumping power,'' generalized to include volume forces.
In previous work the volume forces are actually surface forces 
applied by fixed or flexible elastic obstacles in the channel. These do no net
time-averaged work on the flow since $\mathbf{u} = 0$ on the fixed bodies, and the stored elastic energy remains bounded at large times for moving flexible bodies.   
Our optimization formulation allows an arbitrary body force distribution $\mathbf{f}$, though
we restrict to 2D steady flows and forcing in this work to begin with the simplest case. 
In Appendix \ref{compf} we describe how to compute $\mathbf{f}$ and $p$ from
(\ref{NS}), given $\mathbf{u}$, as the solution to coupled Poisson equations.

\section{Dimensionless equations and boundary conditions\label{sec:Eqns}}

We nondimensionalize lengths by the channel height $H$ and
time by a diffusion time scale $H^2/\kappa$. We have already
nondimensionalized temperature by the temperature of the hot boundary (for
$x \gg \delta$), relative to the temperature of the entering
fluid. Having chosen scales for
length, time and temperature, we need to choose a typical mass scale to nondimensionalize (\ref{Cond}). 
Since mass enters the thermal conductivity, for simplicity we instead chose a thermal conductivity scale to be that 
of the fluid. 

We maximize the dimensionless form of (\ref{Cond}),
\begin{align}
Q = \int_0^{L} \left( \partial_y T |_{y = 1} - \partial_y T |_{y = 0} \right) dx
\end{align}
\nn where $L = L_0/H$, over $T$ satisfying the
dimensionless form of (\ref{AdvDiff}),
\begin{align}
\partial_y \psi \partial_x T - \partial_x \psi \partial_y T - \Delta T &= 0, \label{AdvDiffND}
\end{align}
\nn for a flow $\psi(x,y)$ with rate of viscous dissipation fixed by a constant $Pe^2$,
\begin{align}
\int_0^1 \int_0^L (\Delta \psi)^2 dx dy = Pe^2 = \dot{E} H^2/W \mu \kappa^2. \label{Enstrophy}
\end{align}
\nn Here $W$ is the width of the channel in the out-of-plane direction.
The optimal flow $\psi$ is found by setting to zero the variations of the Lagrangian
\begin{align}
 \mathcal{L} =  \int_0^{L} \left( \partial_y T |_{y = 1} - \partial_y T |_{y = 0} \right) dx + \int_0^1 \int_0^L m(x,y) \left(-\nabla^\perp \psi \cdot \nabla T - \Delta T\right) dx dy + 
\lambda \left(\int_0^1 \int_0^L (\Delta \psi)^2 dx dy - Pe^2 \right). \label{L}
\end{align}
\nn with respect to $T$, $\psi$, and Lagrange multipliers $m$ and $\lambda$ that enforce (\ref{AdvDiffND}) and (\ref{Enstrophy}) respectively. The area integrals
are over the fluid domain, the rectangle in figure \ref{fig:PoiseuilleFlowTempFig}. 
Taking the variations and
integrating by parts, we obtain the following system of three nonlinear partial differential equations (PDEs) plus one integral constraint. We supplement the PDEs with the listed boundary conditions:
\begin{align}
&\mbox{PDE/Constraint}  & &\mbox{Upstream BCs} &  &\mbox{Top BCs}  &  &\mbox{Bottom BCs} & &\mbox{Downstream BCs} \nonumber \\
&0 = \frac{\delta\mathcal{L}}{\delta m} = -\nabla^\perp \psi \cdot \nabla T - \Delta T   & &T=0\; & &T = 1-e^{-\displaystyle\frac{x^2}{\delta^2}} & &T = 1-e^{-\displaystyle\frac{x^2}{\delta^2}} & &\partial_x T = 0 \label{T} \\
&0 = \frac{\delta\mathcal{L}}{\delta T} = \nabla^\perp \psi \cdot \nabla m - \Delta m  & &m=0\; & &m = 1& &m = 1& &\partial_x m + m \partial_y \psi = 0 \label{m} \\
&0 = \frac{\delta\mathcal{L}}{\delta \psi} = -c\nabla^\perp T \cdot \nabla m - 2\Delta^2 \psi  & &\psi = \psi_{top}(3y^2 - 2y^3), & &\psi = \psi_{top},  & &\psi = 0,  &   &\psi = \psi_{top}(3y^2 - 2y^3), \label{psi1} \\
& & &\partial_x \psi = 0 &  &\partial_y \psi = 0  &  &\partial_y \psi = 0 & &\partial_x \psi = 0  \label{psi2}\\
&0 = \frac{\delta\mathcal{L}}{\delta \lambda} = \int_0^1 \int_0^L (\Delta \psi)^2 dx dy - Pe^2. \label{mu}
\end{align}
\nn The boundary conditions for $T$ in (\ref{T}), the usual ones for flow through a heated channel \cite{shoele2014computational}, 
have already been discussed. 

The top and bottom boundary conditions for $\psi$ in (\ref{psi1}) and (\ref{psi2}) correspond to the no-slip condition on the channel walls, with
a net mass flux $\psi_{top}$ (an additional variable to be discussed) through the channel. 
At the upstream and downstream boundaries we set $\psi$ so the flow has a parabolic
velocity profile (Poiseuille flow) with the same mass flux $\psi_{top}$ 
(as required for incompressible flow). If instead any incompressible flow is allowed at these
boundaries, then the ``natural'' boundary conditions are used. These are given by 
setting the boundary terms (not shown)
in $\delta\mathcal{L}/\delta \psi$ to zero. These terms involve $\Delta \psi$ and $\partial_x \Delta \psi$ together
with lower order derivatives of $m$ and $T$. These boundary conditions are not well-posed for the biharmonic operator,
as discussed in \cite{hsiao2008boundary}, and lead to nonconvergence of the finite difference scheme we use.
Furthermore, they do not impose inflow and outflow ($u = \partial_y \psi > 0$) at the corresponding boundaries,
so they may violate the basic physical assumptions of the model. It is possible to use other boundary conditions
at the upstream and downstream boundaries, but an advantage of imposing Poiseuille flow at these boundaries 
is that they allow us to 
generate optimal solutions by continuation from a simple starting solution, Poiseuille flow itself.  
In (\ref{psi1}) we have inserted $c \equiv 1/\lambda$. For the starting Poiseuille flow solution,
$\lambda = \pm\infty$ while $c = 0$, so $c$ is more convenient for computations.

The boundary conditions for $m$, given in 
(\ref{m}), make the boundary terms (not shown) in $\delta\mathcal{L}/\delta m$ equal to zero. 

\section{Solutions\label{sec:Solns}}

Poiseuille flow,
\begin{align}
\psi_{Pois}(x,y) \equiv \psi_{top}(3y^2 - 2y^3),
\end{align}
\nn with a certain value of $\psi_{top}$ and 
certain choices $\{\bar{T}, \bar{m}, \bar{c}\}$ for $\{T, m, c\}$,
respectively, 
gives a convenient starting solution to equations (\ref{T})--(\ref{mu}). 
We can see that it is a solution by first plugging $\psi_{Pois}$ into equation (\ref{mu}). We find that
when $\psi_{top} = \bar{\psi}_{top} \equiv Pe/\sqrt{12 L}$,
the corresponding $\psi_{Pois}$, which we call $\bar{\psi}_{Pois}$, 
satisfies the equation. Second, $\bar{\psi}_{Pois}$ is a biharmonic
function, and it satisfies (\ref{psi1})--(\ref{psi2}) with $c = \bar{c} \equiv 0$ and
$\psi_{top} = \bar{\psi}_{top}$. Plugging $\bar{\psi}_{Pois}$ into (\ref{T}) and
(\ref{m}) and solving, we then obtain $\bar{T}$ and $\bar{m}$,
respectively. 

In fact, this starting solution is the unique solution to (\ref{T})--(\ref{mu})
when $\psi_{top} = Pe/\sqrt{12 L}$. We can show this by writing any solution as
\begin{align}
\psi = \psi_{Pois} + \tilde{\psi}. \label{Decomp}
\end{align}
\nn We will use this decomposition even when $\psi_{top} \neq Pe/\sqrt{12 L}$,
to write solutions in terms of $\tilde{\psi}$.
Plugging (\ref{Decomp}) into (\ref{mu}), we find after integrating by parts and using
homogeneous boundary conditions for $\tilde{\psi}$ that
\begin{align}
12L\psi_{top}^2 + \int_0^1 \int_0^L (\Delta \tilde{\psi})^2 dx dy = Pe^2.   \label{magpsitilde}
\end{align}
\nn If $\psi_{top} = Pe/\sqrt{12 L}$, then $\Delta \tilde{\psi} \equiv 0$,
and again using the homogeneous boundary conditions for $\tilde{\psi}$
implies $\tilde{\psi} \equiv 0$.

\begin{figure}[h]
  \centerline{\includegraphics[width=15cm]
  {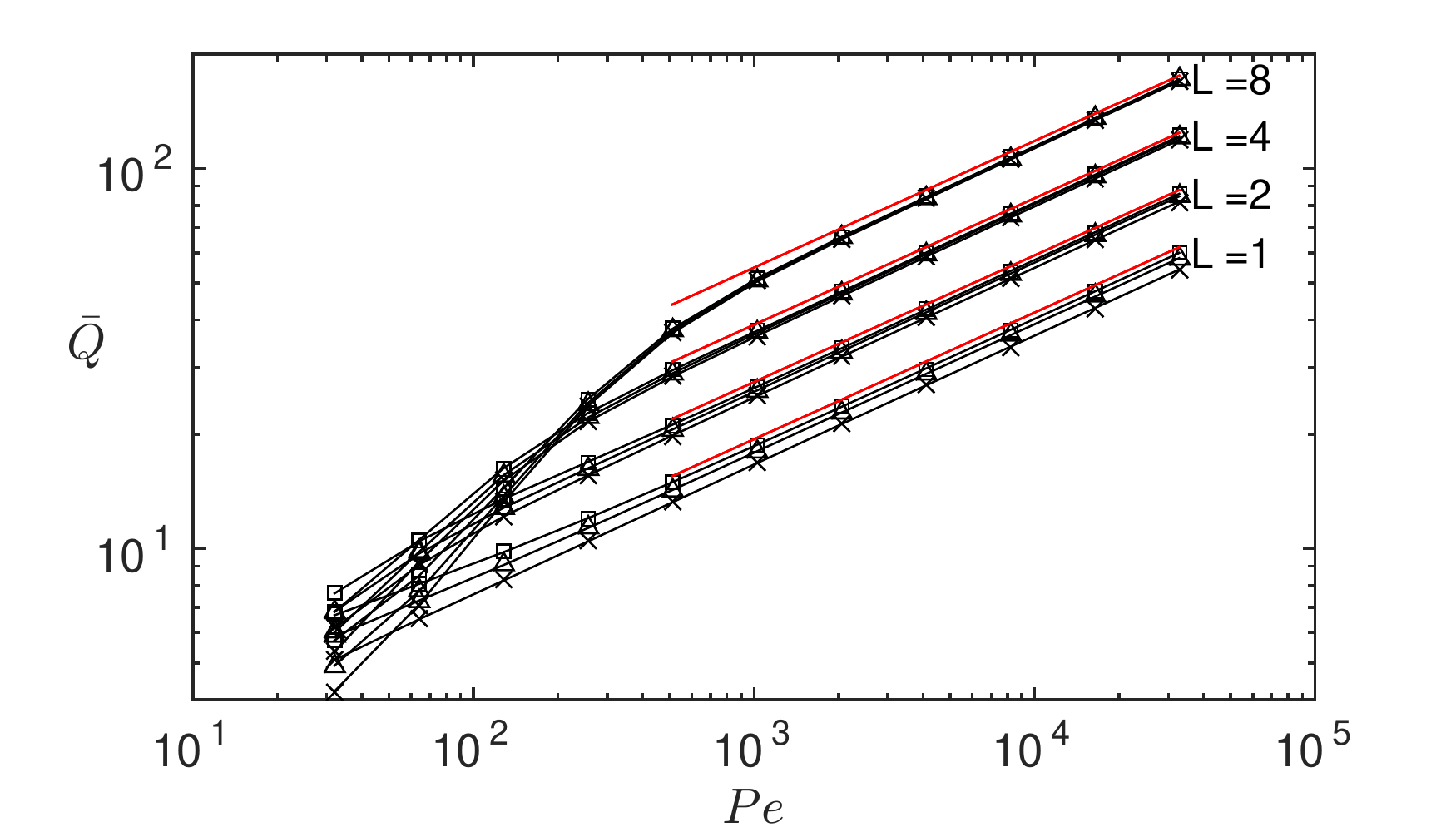}}
  \caption{Heat transfer for Poiseuille flow in a channel, denoted $\bar{Q}$,
at various $Pe$, $L$ = 1, 2, 4, and 8 (labeled), and $\delta$ = 0.05 (squares), 0.1 (triangles),
and 0.2 (crosses). 
The red lines show, for each $L$, the leading-order term in $Q$ from the 
similarity solution of \cite{worsoe1967heat} (based on
that of \cite{leveque1928laws} and reported on page 173 of 
\cite{shah2014laminar}).}
\label{fig:QPoiseuilleFig}
\end{figure}

The benchmark solution at a given $Pe$ is $\bar{\psi}_{Pois}$, so we first show how $Q$ for this solution, denoted $\bar{Q}$, varies with parameters
 before giving the results for other optima. Figure \ref{fig:QPoiseuilleFig} plots
$\bar{Q}$ versus $Pe$ for four different values of $L$. For each $L$, we have a cluster of
three lines (labeled by the value of $L$) giving $Q$ values at three $\delta$, 0.05, 0.1, and 0.2, to show the effect of this smoothing
parameter. We find a slight increase in the heat transferred as $\delta \to 0$.
The red line gives the leading term for the similarity solution of \cite{worsoe1967heat} (from 
\cite{shah2014laminar}), which corresponds to the case $\delta = 0$. Here 
the temperature profile is a function of a similarity variable
\begin{align} 
\eta \equiv y Pe^{1/3}/x^{1/3} L^{1/6} \label{eta}
\end{align}
near the bottom wall (with a symmetrically reflected boundary layer at the top wall,
taking $1-y$ in place of $y$). The combination $Pe^{1/3}/L^{1/6}$ is
essentially flow speed to the 1/3-power, since for Poiseuille flow $\bar{\psi}_{top} \equiv Pe/\sqrt{12 L}$.  The width of the boundary layer in the $y$ direction
at a given $x$ scales as $Pe^{-1/3}L^{1/6}$ and the temperature
contours in Figure \ref{fig:PoiseuilleFlowTempFig} grow like a cube root of
$x$ moving downstream.
The total heat transferred for the similarity solution is given
by \cite{worsoe1967heat}, and in terms of our 
parameters ($Pe$ and $L$), it is 
$\bar{Q} \sim 1.85 (4/3)^{1/6} Pe^{1/3} L^{1/2}$, shown by the red lines
in Figure \ref{fig:QPoiseuilleFig}. At smaller $Pe$ (slower flows) and 
larger $L$ (longer channels), the values deviate from the red lines because
the thermal boundary layers on the top and bottom
(shown in Figure \ref{fig:PoiseuilleFlowTempFig}) are large enough 
to intersect within the channel, and the solutions are no longer
well-approximated by the similarity solution.

Our procedure for generating optimal flows is then the following continuation scheme. 
For each $Pe$, we start with $\psi_{top} = \bar{\psi}_{top} = Pe/\sqrt{12 L}$ and the unique optimum
$\{\bar{T}, \bar{m}, \bar{\psi}_{Pois}, \bar{c}\}$. We then decrease 
$\psi_{top}$ gradually, and compute the solution to (\ref{T})--(\ref{mu}) 
using Newton's method
with the solution at the previous value of $\psi_{top}$ as an initial
guess. The best among these solutions is not necessarily
a global optimum. For example, there could be optima that
are not smoothly continued from the starting solution. However, 
from the computed solutions we do
obtain a local optimum which provides a lower bound on the performance of a
global optimum. Furthermore, because there is a unique starting solution,
we can characterize the set of possible optima in the vicinity of the
starting solution (discussed more fully in the next section, \ref{sec:Approx}).

\begin{figure}[h]
  \centerline{\includegraphics[width=15cm]
  {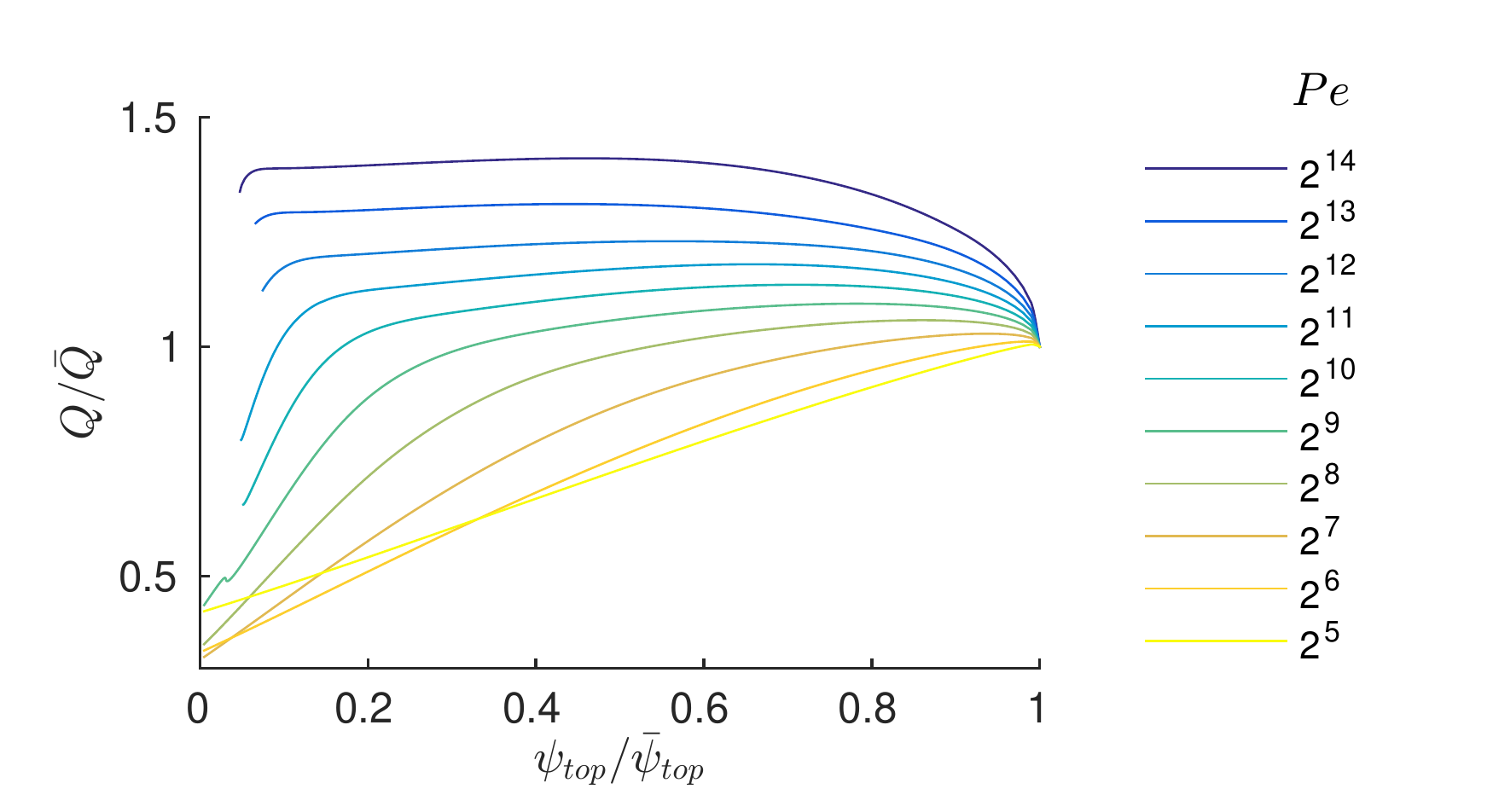}}
  \caption{Values of heat transferred $Q$, relative to that of the starting Poiseuille
solution $\bar{Q}$, for optimal flows with $Pe$ ranging from $2^5$ to $2^{14}$ and $\psi_{top}/\bar{\psi}_{top}$ ranging from 1 down to 0-0.1 (where our Newton's method stops converging). The channel length $L$ = 2.}
\label{fig:QvsPeFig}
\end{figure}

In Figure \ref{fig:QvsPeFig} we show the heat transferred for the optimal solutions
as the continuation parameter $\psi_{top}/\bar{\psi}_{top}$ ranges from 1 down to 0--0.1 (where our Newton's method stops converging). 

For the smallest $Pe$ = 32 the
optimum occurs for $\psi_{top}/\bar{\psi}_{top}$ very close to 1, so Poiseuille
flow is almost optimal. As the flow 
becomes even slower ($u \sim Pe/\sqrt{L} \lesssim 1$) or the channel longer (larger $L$), the temperature is nearly its maximum (unity) 
all along the outflow boundary, so the heat flux out of this boundary
is almost the same as the net flow rate: 
\begin{align}
\int_0^1 T\mathbf{u} \cdot \mathbf{n}|_{x = L}  \,dy \approx
\int_0^1 1 \,\mathbf{u} \cdot \mathbf{n}|_{x = L}  \,dy = \psi_{top}.
\end{align}
\nn Since Poiseuille flow maximizes the flow rate at a given $Pe$, it
is not surprising that it is nearly optimal. More precisely, if we consider 
the similarity solution for temperature in a Poiseuille flow mentioned above,
the temperature is close to unity throughout the outflow boundary if 
the boundary layer extends well past the middle of the 
channel ($y = 1/2$) at $x = L$, which means the similarity variable is
$\ll 1$ there:
$1 \gg \eta \equiv y Pe^{1/3}/x^{1/3} L^{1/6} \sim Pe^{1/3}/L^{1/2}$
at $y = 1/2$ and $x = L$.
In the example above, $Pe = 32$ and $L = 2$, so $Pe^{1/3}/L^{1/2} = 2.2$ is not yet
in the regime $Pe^{1/3}/L^{1/2} \ll 1$, but is close to it.
In this regime, the optimal flow is very close to Poiseuille flow. 
As $\bar{\psi}_{top} \equiv Pe/\sqrt{12 L}$ approaches zero, the upstream and downstream temperature boundary conditions are no longer physically realistic since they assume  
strong advection ($Pe/\sqrt{L} \gg 1$). 

\begin{figure}
  \centerline{\includegraphics[width=18cm]
  {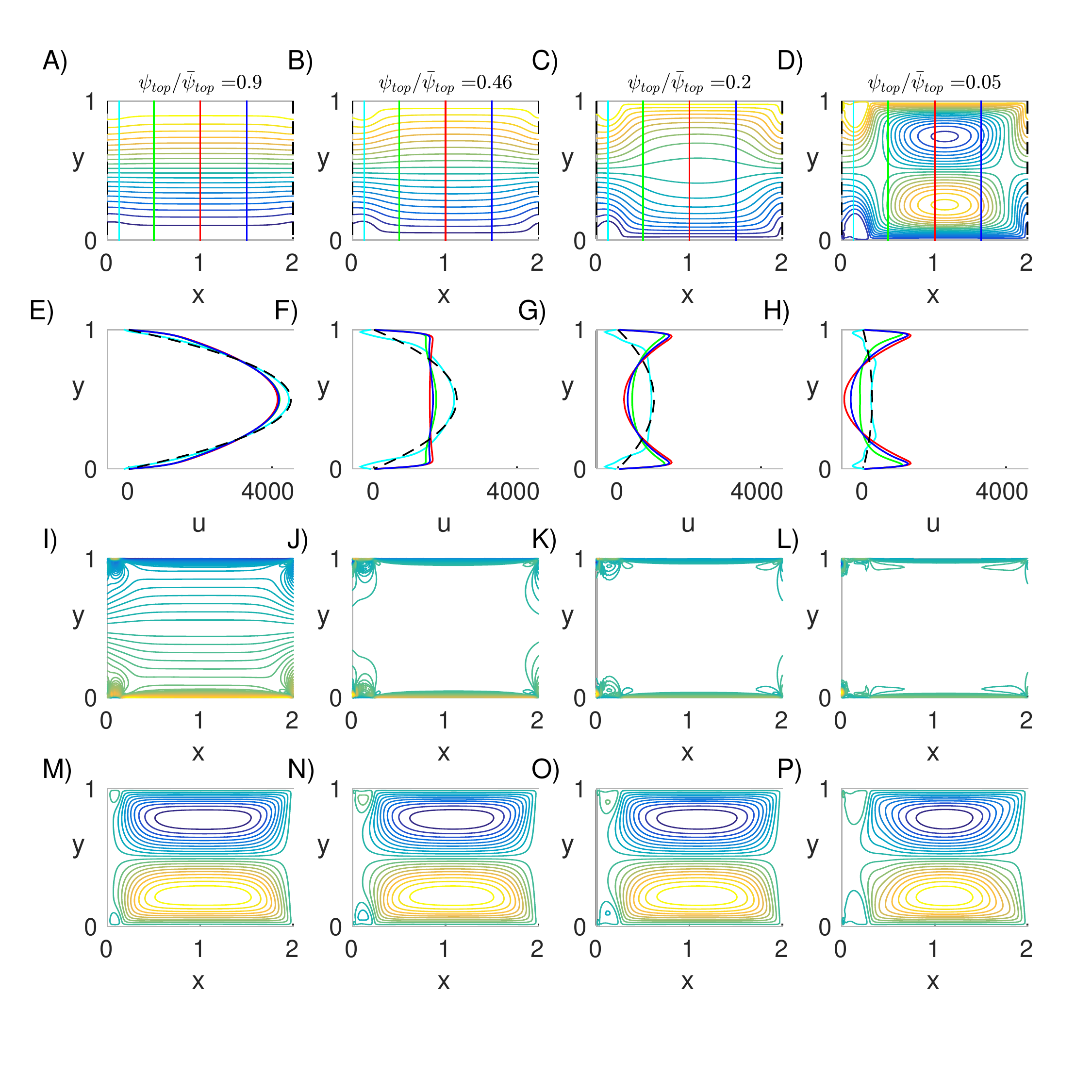}}
  \caption{Optimal flows for $Pe = 2^{14}$ and four different
values of $\psi_{top}/\bar{\psi}_{top}$: 0.9 (first column), 0.46 (second column),
0.2 (third column), and 0.05 (fourth column). The values of $Q$ are: 84.4 (first column), 95.2 (second column), 94.1 (third column), and 91.3 (fourth column).
The values of $\tilde{\psi}_{max}$ are (I) 69.5, (J) 141.2, (K) 156.7, and (L) 149.3.
 }
\label{fig:FlowVsaFig}
\end{figure}

As we move from $Pe = 32$ to larger $Pe$ in figure \ref{fig:QvsPeFig}, we
see that the $Q$-maximizing flow occurs at a decreasing value
of $\psi_{top}/\bar{\psi}_{top}$, and moves further from Poiseuille flow. 
Also, the heat transfer improvement
$Q/\bar{Q}$ increases with increasing $Pe$. To understand the corresponding
physics, we focus on the highest $Pe$ shown in figure \ref{fig:QvsPeFig}: $2^{14}$.
At this $Pe$ we plot the flows at four values of $\psi_{top}/\bar{\psi}_{top}$, in the four columns on figure \ref{fig:FlowVsaFig}. Moving from the first column
to the fourth column, we move away from the Poiseuille flow optimum by
decreasing the flux $\psi_{top}/\bar{\psi}_{top}$ to 0.9, 0.46, 0.2, and
0.05. The different rows plot different flow quantities for each optimum. 
The first row shows streamlines for the total flow, $\psi$. The
second row shows the horizontal velocity profile along six cross-sections
marked in the first row (with corresponding colors). The third row shows
contours of the vorticity fields. The fourth row shows streamlines for
$\tilde{\psi}$, the total flow minus the Poiseuille flow component,
defined in (\ref{Decomp}). 

In the first column, the flow is close to the
initial Poiseuille flow optimum. The streamlines (A) are nearly unidirectional,
with bending slightly towards the wall at the inflow and away from the wall at the outflow. The flow profiles $u(y)$ (E) are prescribed as Poiseuille flow 
(black dashed lines) at the inflow
and outflow boundaries. The profiles at 1/4 (green), 1/2 (red),
and 3/4 (blue) of the channel length are nearly identical and are
perturbations of the Poiseuille inflow
and outflow, with a slight weakening of the parabolic flow in the center and
a strengthening of the flow near the walls. The vorticity (I) has a nearly
 uniform gradient (as for Poiseuille flow) around $y = 1/2$, but with
a sharp increase near the boundaries. The $\tilde{\psi}$-component (M)
shows two large vortices, equal and opposite, with downstream flow
near the walls and upstream flow near $y = 1/2$. A smaller pair of
oppositely-signed vortices is located on the wall near the inflow boundary,
where the wall temperature changes from 0 to 1.

The second column shows the flow that maximizes $Q$ over all
$\psi_{top}/\bar{\psi}_{top}$ at this $Pe$. Here $Q$ is 41\% higher
than that of the initial Poiseuille flow, while the net flow through 
the channel is reduced to 46\% of its initial value. Panel B shows streamlines
that are nearly unidirectional as in A, but spread towards the walls.
The spacing between the streamlines is nearly uniform, indicating a nearly
uniform flow. This is shown clearly in panel F, where the green, red, and
blue velocity profiles are all nearly uniform except for sharp boundary layers
near the walls. The vorticity is now almost entirely at the walls (panel J),
with additional vorticity near the inflow and outflow boundaries at the walls.
The $\tilde{\psi}$-component (N) has a spatial distribution similar to 
that in first column, with somewhat larger vortices at the corners near the
inflow boundary. In the third column, the net flow is only 0.2 times its
initial value, but $Q$ is still 99\% that of the second column. The flow is now
very nonuniform near $y = 1/2$ (panel C). Panel G shows that it resembles 
a backward facing parabolic profile. Otherwise, the features are similar to
those in the second column. In the fourth column, the net flow is reduced to
0.05 its initial value. The flow divides into
regions of downstream flow near the walls and two closed eddies with upstream
flow in the center of the channel near $y = 1/2$ and $1/2 < x < 3/2$.
However, the flow profiles (H) are not so different from the third column
(G). Again we have backward facing parabolic profiles for $1/2 < x < 3/2$,
which cross zero, corresponding to upstream flow. Surprisingly, the 
$\tilde{\psi}$-component (P) is not much different now, and 
$Q$ is almost (96\%) that of the optimum (second column).

We would like to explain some of the key features of the optimal
flows, and in particular, 
determine for these flows how $Q$ scales with $Pe$ and $L$. We will
use a combination of approximations, computations, and asymptotics.
First we will give a simple intuitive explanation for what happens
as $\psi_{top}/\bar{\psi}_{top}$ is decreased from 1 at a given $Pe$.
For a Poiseuille flow with large $Pe$, the temperature is large 
near the walls and essentially zero elsewhere (as in figure \ref{fig:PoiseuilleFlowTempFig}, except that in figure \ref{fig:FlowVsaFig} $Pe$ is larger ($2^{14}$), 
so the temperature boundary layer is a factor of $2^{4/3}\approx 2.5$ thinner in
Poiseuille flow). To move
more heat out of the channel, it is advantageous to increase the downstream
flow where the temperature is nonzero (near the walls), which means having
a steeper velocity gradient near the no-slip walls. To keep the total
enstrophy ($Pe^2$) constant, enstrophy is then depleted from the center of the channel
(near $y = 1/2$), slowing the downstream flow there. 
Since the temperature is zero there for all $x$, this depletion does not decrease the
heat flux through the channel. The net effect of moving enstrophy from
the channel center to the walls is to increase the heat flux out. The
second column of figure \ref{fig:FlowVsaFig} is optimal because
the flow is uniform in the center of the channel, so no enstrophy is
spent there. All of the enstrophy is in the boundary layers near the walls,
giving the maximum possible downstream flow there. In the third and fourth
columns, with much smaller net flux prescribed, some enstrophy is put back into the center of the channel, which takes enstrophy from the wall boundary layers and therefore slows them down. However, the velocity gradient is still large enough
near the walls to transfer nearly the optimal amount of heat out of the channel. 
There may be some additional compensating 
effects for the slower flows, in that there is more
time to mix the hotter fluid near the wall with colder fluid away from the wall
before the fluid leaves the channel, as proposed by \cite{gerty2008fluidic,hidalgo2010heat,shoele2014computational,jha2015small}. However, unlike the
heat transfer enhancement strategies using vortex shedding, here the optimal flow
(second column) is mainly in the $x$-direction, and does not contain 
a sequence of coherent vortices. The vortices near the inflow boundary
are the only occurrence of discrete vortices in the flow. 

In the last paragraph we have given the physical intuition for the optimal flow structure. In the next section, we analyze the mathematical structure and asymptotic scalings using a decoupled
approximation for the optimal flows. Readers who are less interested in the mathematical
details may wish to skip to the following section, which explains the asymptotic scalings  for the fully coupled system using a unidirectional flow approximation, and gives
physical interpretations.

\section{Decoupled approximation\label{sec:Approx}}

In order to determine how the heat transfer $Q$ scales with $Pe$ and $L$, we use the
method of successive approximations \cite{lin1988mathematics} for the nonlinear equations (\ref{T})--(\ref{mu}).
When $\psi_{top}$ is close to $\bar{\psi}_{top}$, $\psi \approx 
\bar{\psi}_{Pois} \approx \psi_{Pois}$, $\|\tilde{\psi}\| \ll \|\psi_{Pois}\|$,
$T \approx \bar{T}$, and $m \approx \bar{m}$.
This motivates the following approximation to 
(\ref{T})--(\ref{mu}). First we write the
$\psi$ equations (\ref{psi1})--(\ref{psi2}) in terms of $\tilde{\psi}$ defined
in (\ref{Decomp})
(no approximation yet). Then we approximate $T$ and $m$ in (\ref{psi1})--(\ref{psi2}) by their values in the initial Poiseuille solution, $\bar{T}$ and $\bar{m}$, respectively. Then the coupling
from $\bar{m}$ and $\bar{T}$ to $\tilde{\psi}$ is one-way:


\begin{align}
&\mbox{PDE/Constraint}  & &\mbox{Upstream BCs} &  &\mbox{Top BCs}  &  &\mbox{Bottom BCs} & &\mbox{Downstream BCs} \nonumber \\
&0  = -\nabla^\perp \bar{\psi}_{Pois} \cdot \nabla \bar{T} - \Delta \bar{T}   & &\bar{T}=0\; & &\bar{T} = 1-e^{-\displaystyle\frac{x^2}{\delta^2}} & &\bar{T} = 1-e^{-\displaystyle\frac{x^2}{\delta^2}} & &\partial_x \bar{T} = 0 \label{barT} \\
&0  = \nabla^\perp \bar{\psi}_{Pois} \cdot \nabla \bar{m} - \Delta \bar{m}  & &\bar{m}=0\; & &\bar{m} = 1& &\bar{m} = 1& &\partial_x \bar{m} + \bar{m} \partial_y \bar{\psi}_{Pois} = 0 \label{barm} \\
&0  = -c_1\nabla^\perp \bar{T} \cdot \nabla \bar{m} - 2\Delta^2 \tilde{\psi}_1  & &\tilde{\psi}_1 = 0, & &\tilde{\psi}_1 = 0,  & &\tilde{\psi}_1 = 0,  &   &\tilde{\psi}_1 = 0, \label{psi1tilde} \\
& & &\partial_x \tilde{\psi}_1 = 0 &  &\partial_y \tilde{\psi}_1 = 0  &  &\partial_y \tilde{\psi}_1 = 0 & &\partial_x \tilde{\psi}_1 = 0  \label{psi2tilde}\\
&0 = \int_0^1 \int_0^L (\Delta \tilde{\psi}_1)^2 dx dy + 12L\psi_{top}^2 - Pe^2. \label{mutilde} \\
&0  = -\nabla^\perp \left({\psi}_{Pois}+\tilde{\psi}_1\right) \cdot \nabla T_1 - \Delta T_1   & &T_1=0\; & &T_1 = 1-e^{-\displaystyle\frac{x^2}{\delta^2}} & &T_1 = 1-e^{-\displaystyle\frac{x^2}{\delta^2}} & &\partial_x T_1 = 0 \label{T1}
\end{align}
\nn Note that $\tilde{\psi}_1$ in (\ref{psi1tilde})--(\ref{psi2tilde}) is
an approximation to $\tilde{\psi}$ (since $\bar{T}$ and $\bar{m}$ are computed
using $\bar{\psi}_{Pois}$ instead of $\psi = \psi_{Pois} + \tilde{\psi}$).
Equation (\ref{T1}) computes the temperature field $T_1$ for this approximation
to the optimal flow, from which we compute the heat transferred, $Q_1$.

In (\ref{psi1tilde}) $c_1$ is an approximation to $c$, 
computed by plugging $\tilde{\psi}_1$ from (\ref{psi1tilde}) into (\ref{mutilde}):
\begin{align}
c_1 =  \pm2\sqrt{Pe^2 - 12L \psi_{top}^2} \Bigg/
\sqrt{\int_0^1 \int_0^L \left(\Delta^{-1} (\nabla^\perp \bar{T} \cdot \nabla \bar{m})\right)^2 dx dy}. \label{ctilde}
\end{align}
\nn Due to the $\pm$ in (\ref{ctilde}), there are two branches of approximate
optimal flows
\begin{align}
\tilde{\psi}_1 = \mp\Delta^{-2} (\nabla^\perp \bar{T} \cdot \nabla \bar{m})\sqrt{Pe^2 - 12L \psi_{top}^2} \Bigg/
\sqrt{\int_0^1 \int_0^L \left(\Delta^{-1} (\nabla^\perp \bar{T} \cdot \nabla \bar{m})\right)^2 dx dy} \label{psi1tildesoln}
\end{align}
leading away from the starting optimum. 
The negative sign gives a local minimizer of heat transferred $Q_1$ and the positive sign
gives a local maximizer, so this is the branch we compute. The approximation
we have made should be good for 
$\psi_{top}$ close to $\bar{\psi}_{top}$ but the fourth row of 
figure \ref{fig:FlowVsaFig} gives us hope that it works well for a much
larger range of $\psi_{top}$, because there is relatively little change
in the spatial distribution of $\tilde{\psi}$ for 
$0.05 \leq \psi_{top}/\bar{\psi}_{top} \leq 0.9$. As $\psi_{top}$ varies,
$\tilde{\psi}_1$ in (\ref{psi1tildesoln}) has a fixed
spatial distribution, scaled by different constants
$\sqrt{Pe^2 - 12L \psi_{top}^2}$.

\begin{figure}
  \centerline{\includegraphics[width=15cm]
  {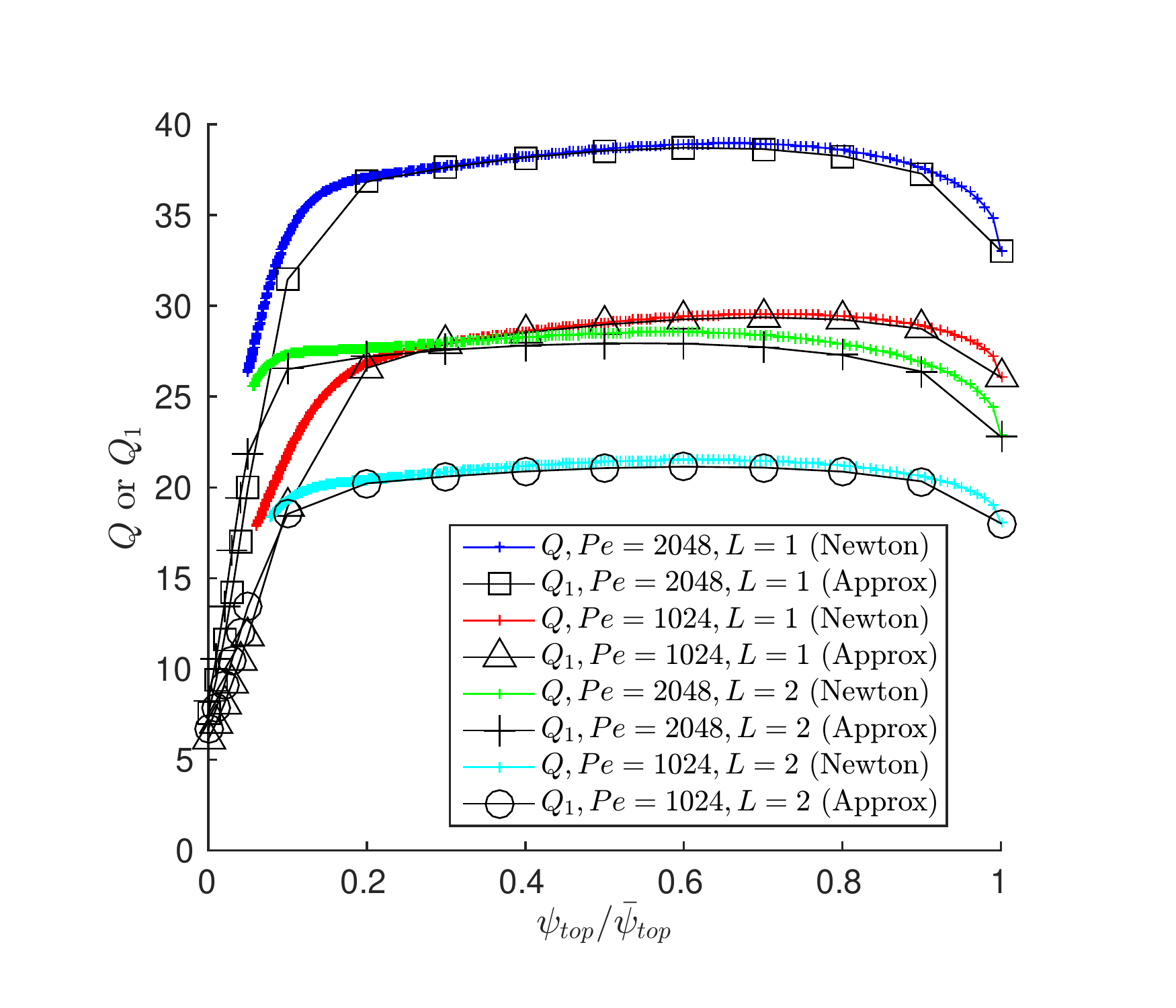}}
  \caption{Comparison of the heat transferred by the optimal flows ($Q$, colored lines) computed by solving the nonlinear equations (\ref{T})--(\ref{mu}) with
that transferred by the approximate optimal flows ($Q_1$, black lines), computed by solving  (\ref{barT})--(\ref{T1}). Two channel lengths ($L = 1, 2$) and
two values of $Pe$ (1024 and 2048) are used.}
\label{fig:ApproxNewtonFig}
\end{figure}

In figure \ref{fig:ApproxNewtonFig} we compare the heat transferred
by the approximate optimal flow $\psi_1 \equiv {\psi}_{Pois}+\tilde{\psi}_1$ (using (\ref{T1}))
with that from $\psi = {\psi}_{Pois}+\tilde{\psi}$, the flow solution to the full
nonlinear equations (\ref{T})--(\ref{mu}). We use two channel lengths $L = 1, 2$ and
two values of $Pe$, 1024 and 2048. For each of the four parameter combinations
we plot the heat transferred versus $\psi_{top}/\bar{\psi}_{top}$ for the nonlinear solution
($Q$, colored lines) and the approximation ($Q_1$, black lines). The agreement is
remarkably good over a broad range, 
$0.1 \leq \psi_{top}/\bar{\psi}_{top} \leq 1$. In the approximate flow
equation (\ref{psi1tilde}) we used $\bar{T}$, the temperature
due to the starting Poiseuille flow $\bar{\psi}_{Pois}$, instead of $T$, the temperature
due to the actual flows $\psi$, which are quite different from Poiseuille
flow, especially for smaller $\psi_{top}/\bar{\psi}_{top}$, as shown
by the first row of figure \ref{fig:FlowVsaFig}. 

Although 
$\bar{\psi}_{Pois}$ is quite different from $\psi$, it turns out
that $\bar{T}$ is similar to $T$. The reason, briefly, is that
$\bar{T}$ and $T$ both transition from 1 to 0 
over thin boundary layers near the wall, so they are only
affected by the flows there. In the boundary layers, 
$\psi$ and $\bar{\psi}_{Pois}$
both have sharp linear flow gradients near the wall, with
that from $\psi$ somewhat sharper. The differences between
$\psi$ and $\bar{\psi}_{Pois}$ outside the boundary layers
have little effect on the temperature field because it is nearly zero 
there.

\begin{figure}
  \centerline{\includegraphics[width=15cm]
  {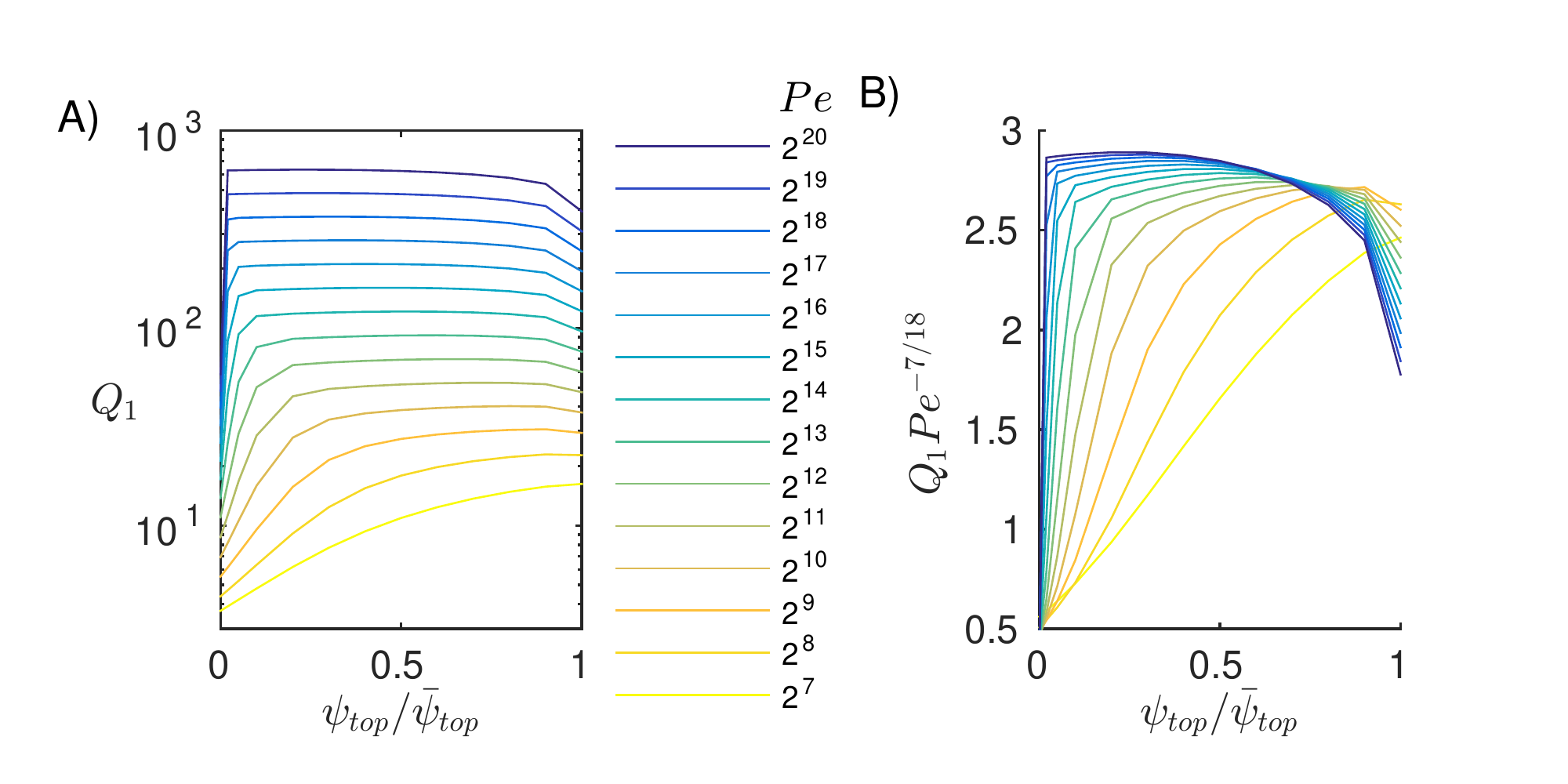}}
  \caption{Heat transferred ($Q_1$) by the approximate
optimal flows. A) $Q_1$ versus $\psi_{top}/\bar{\psi}_{top}$ at several
values of $Pe$ (listed to the right). B) $Q_1$ rescaled by $Pe^{7/18}$.}
\label{fig:QApproxFig}
\end{figure}

As $Pe$ grows, it becomes 
difficult to obtain convergence with Newton's method
for our finite-difference discretization of (\ref{T})--(\ref{mu}). The
Jacobian matrix contains a biharmonic operator, for which
the condition number scales as grid spacing to the $-4$ power.
At large $Pe$, a fine grid is needed near the boundaries to resolve
the boundary layers in $T$, $m$, and $\psi$. To obtain the asymptotic scaling
of $Q$ with $Pe$ at larger $Pe$ than the maximum 
value in figure \ref{fig:QvsPeFig} ($= 2^{14}$), we proceed with
calculations of the approximate optimal flow 
$\psi_1 \equiv {\psi}_{Pois}+\tilde{\psi}_1$ and
the resulting temperature field $T_1$ in (\ref{psi1tilde}), (\ref{psi2tilde}),
and (\ref{T1}), respectively. These can be calculated using
direct solvers for the linear systems in (\ref{barT})--(\ref{T1}),
so we avoid Newton's method and associated convergence issues.
Although the approximate flows do not achieve the
optimal heat transfer, their heat transfer is close to that of 
the optimal flows in figure \ref{fig:ApproxNewtonFig}, and
they show us how to obtain the right scalings for the optimal flows.

In figure \ref{fig:QApproxFig}A we plot $Q_1$, the heat transferred by the
approximate optimal flows $\psi_1$, up to $Pe = 2^{20}$, and
across the full range $0 \leq \psi_{top}/\bar{\psi}_{top} \leq 1$.
When $\psi_{top}/\bar{\psi}_{top} = 1$ the solution is
$\bar{\psi}_{Pois}$ with heat transfer $\bar{Q} \sim Pe^{1/3}$.
The maximum of $Q_1$ over $\psi_{top}/\bar{\psi}_{top}$
exceeds $\bar{Q}$, and the difference grows with $Pe$.
We will subsequently argue that max($Q_1$) $\sim Pe^{7/18}$.
In panel B we plot $Q_1$ divided by $Pe^{7/18}$ and
see an approximate collapse of the maxima of the curves.
The slight improvement of 7/18 over 1/3 is due to 
reconfiguring the flow as $Pe$ increases 
rather than simply speeding up a given (Poiseuille) flow
profile.

\begin{figure}
  \centerline{\includegraphics[width=17cm]
  {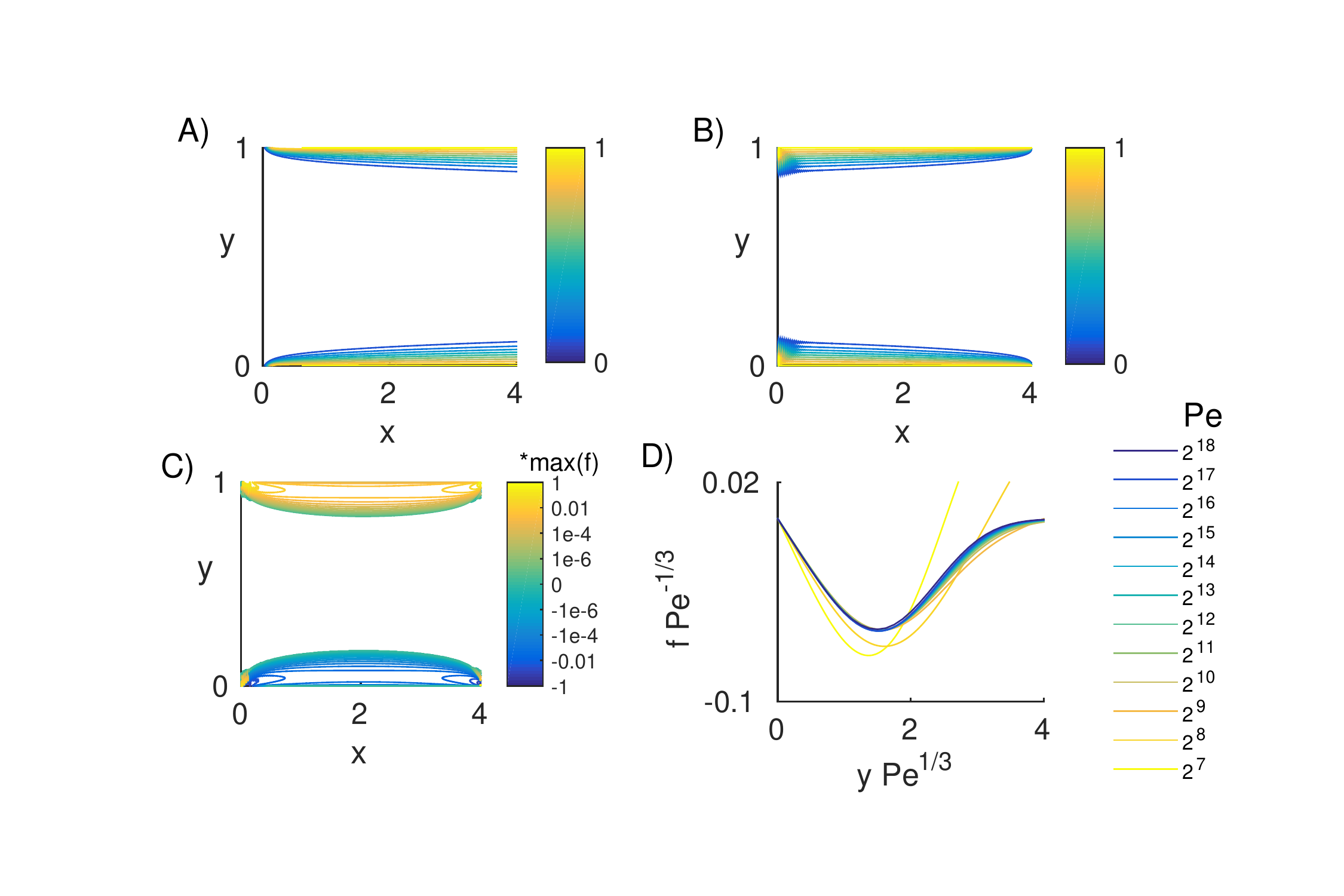}}
  \caption{Plots of the components of the source terms for the
approximate optimal flow, $\tilde{\psi}_1$ in (\ref{psi1tilde}).
A: Contours of $\bar{T}$
at $Pe = 2^{15}$, $L = 4$, and $\delta = 0.05$. 
B: Contours of $\bar{m}$ at the same parameters.
C: Contours of $f \equiv \partial_x \bar{T} \partial_y \bar{m} - \partial_y \bar{T} \partial_x \bar{m}$ at the same parameters. D: Profiles of
$f$ along the channel midline at $x = L/2 = 2$, at various $Pe$, 
rescaled near $y = 0$ to show the asymptotic behavior.}
\label{fig:fApproxFig}
\end{figure}

To understand how a $Pe^{7/18}$ scaling arises from the equations,
we plot in figure \ref{fig:fApproxFig} the contributions to 
the ``source term'' in the biharmonic equation for $\tilde{\psi}_1$ (\ref{psi1tilde}). In panel A, we plot the contours of $\bar{T}$
at $Pe = 2^{15}$ and $L = 4$, showing the same kind of Poiseuille-flow
temperature boundary layer as in figure 
\ref{fig:PoiseuilleFlowTempFig}, but narrower at this larger $Pe$.
Panel B shows the corresponding $\bar{m}$. Since $\bar{m}$ solves
the backward advection diffusion equation (\ref{barm}), we have a
profile which is almost that of $\bar{T}$ reflected in the 
$x$-midpoint of the channel. The boundary conditions in (\ref{barm}) are
somewhat different than in (\ref{barT}). Along the walls
$\bar{m} = 1$, close to the values of $\bar{T}$ since
$\delta$ is small (0.05 here). At $x = L$, 
$\partial_x \bar{m} + \bar{m} \partial_y \bar{\psi}_{Pois} = 0$,
and since $\partial_x \bar{m}$ is exponentially small outside
the boundary layer, but $\partial_y \bar{\psi}_{Pois}$ is $O(Pe) \gg 1$, $\bar{m}$ is also exponentially small outside
the boundary layer. At $x = 0$, $\bar{m} = 0$,
but since we have strong advection in the $-x$ direction, this
only affects the solution in a thin layer near $x = 0$, too thin
to be visible in panel B. In panel C we plot on a logarithmic scale
contours of
$f \equiv \partial_x \bar{T} \partial_y \bar{m} - \partial_y \bar{T} \partial_x \bar{m}$, 
the source term
for $\tilde{\psi}_1$ in (\ref{psi1tilde}), up to a constant ($c_1$).
Because $\bar{T}$ and $\bar{m}$ are essentially constant (zero) 
outside a distance $\sim Pe^{-1/3}$ from the walls, so is 
the source term. Using the approximate 
similarity solutions $\bar{T} = g(\eta)$, $\bar{m} = g(\tilde{\eta})$,
with $\eta$ in (\ref{eta}) and
\begin{align} 
\tilde{\eta} \equiv y Pe^{1/3}/(L-x)^{1/3} L^{1/6}, \label{etat}
\end{align}
\nn we have
\begin{align} 
f = -\frac{Pe^{1/3}}{3 L^{1/6}} \left(\frac{\eta}{x(L-x)^{1/3}} 
+ \frac{\tilde{\eta}}{x^{1/3}(L-x)}\right)g'(\eta)g'(\tilde{\eta}). \label{f}
\end{align}
\nn At a given $x$, $f \sim Pe^{1/3}$ in boundary layers of width
$\sim Pe^{-1/3}$ in $y$. Panel D shows this behavior in the
bottom boundary layer for
$x = L/2$ ($= 2$ here), with a collapse of the curves at larger $Pe$.
Panel C shows an example of how 
$f$ varies gradually in $x$ over 
the middle half of the channel ($L/4 \leq x \leq 3L/4$).

To understand what kind of flow this $f$ produces, we define
$\tilde{\psi}_0 \equiv \tilde{\psi}_1/c_1$, so by
(\ref{psi1tilde})--(\ref{psi2tilde}),
\begin{align}
&\mbox{PDE}  & &\mbox{Upstream BCs} &  &\mbox{Top BCs}  &  &\mbox{Bottom BCs} & &\mbox{Downstream BCs} \nonumber \\
&- 2\Delta^2 \tilde{\psi}_0 = f,  & &\tilde{\psi}_0 = 0, & &\tilde{\psi}_0 = 0,  & &\tilde{\psi}_0 = 0,  &   &\tilde{\psi}_0 = 0, \label{psi01tilde} \\
& & &\partial_x \tilde{\psi}_0 = 0 &  &\partial_y \tilde{\psi}_0 = 0  &  &\partial_y \tilde{\psi}_0 = 0 & &\partial_x \tilde{\psi}_0 = 0  \label{psi02tilde}
\end{align}
\nn $\tilde{\psi}_0$ is simply $\tilde{\psi}_1$ without the as-yet-unknown
constant $c_1$. The equation for $\tilde{\psi}_0$ is that for
a thin rectangular plate with unit bending modulus clamped on all sides 
loaded by a force per unit area $-f/2$. By (\ref{f}), $f$ is
zero outside of the boundary layers at the $y$ boundaries ($g'$ tends to
zero much more rapidly than $\eta$ and $\tilde{\eta}$ grow, moving
away from the boundaries). Also, $f$ has equal magnitude and
opposite sign at corresponding points in the two boundary layers
at $y = 0$ and 1.
In the boundary layers, $y$-derivatives of $f$ are larger than $x$-derivatives by a factor involving $Pe^{1/3}$. By the biharmonic
equation in (\ref{psi01tilde}),
we expect similar boundary layers to appear in $\tilde{\psi}_0$ and
its derivatives. In particular, we expect 
$|\partial_y^4 \tilde{\psi}_0| \gg |\partial_x^2 \partial_y^2 \tilde{\psi}_0|,
|\partial_x^4 \tilde{\psi}_0|$. Outside the boundary layers,
$f \approx 0$, so 
$\partial_y^4 \tilde{\psi}_0 \approx -2\partial_x^2 \partial_y^2 \tilde{\psi}_0 -\partial_x^4 \tilde{\psi}_0$, much smaller
than $\partial_y^4 \tilde{\psi}_0$ inside the boundary layer.
Away from the upstream and downstream boundaries, the $x$-derivatives
are expected to be small since $f$ varies slowly with $x$ in this region
and the top and bottom boundary conditions are $x$-independent. Also,
the domain is generally narrower in $y$ than in $x$, and the
$x$ boundaries have a limited effect in the middle part of the domain
for such problems
(elliptic PDEs in long thin domains \cite{howison2005practical}).

\begin{figure}
  \centerline{\includegraphics[width=17cm]
  {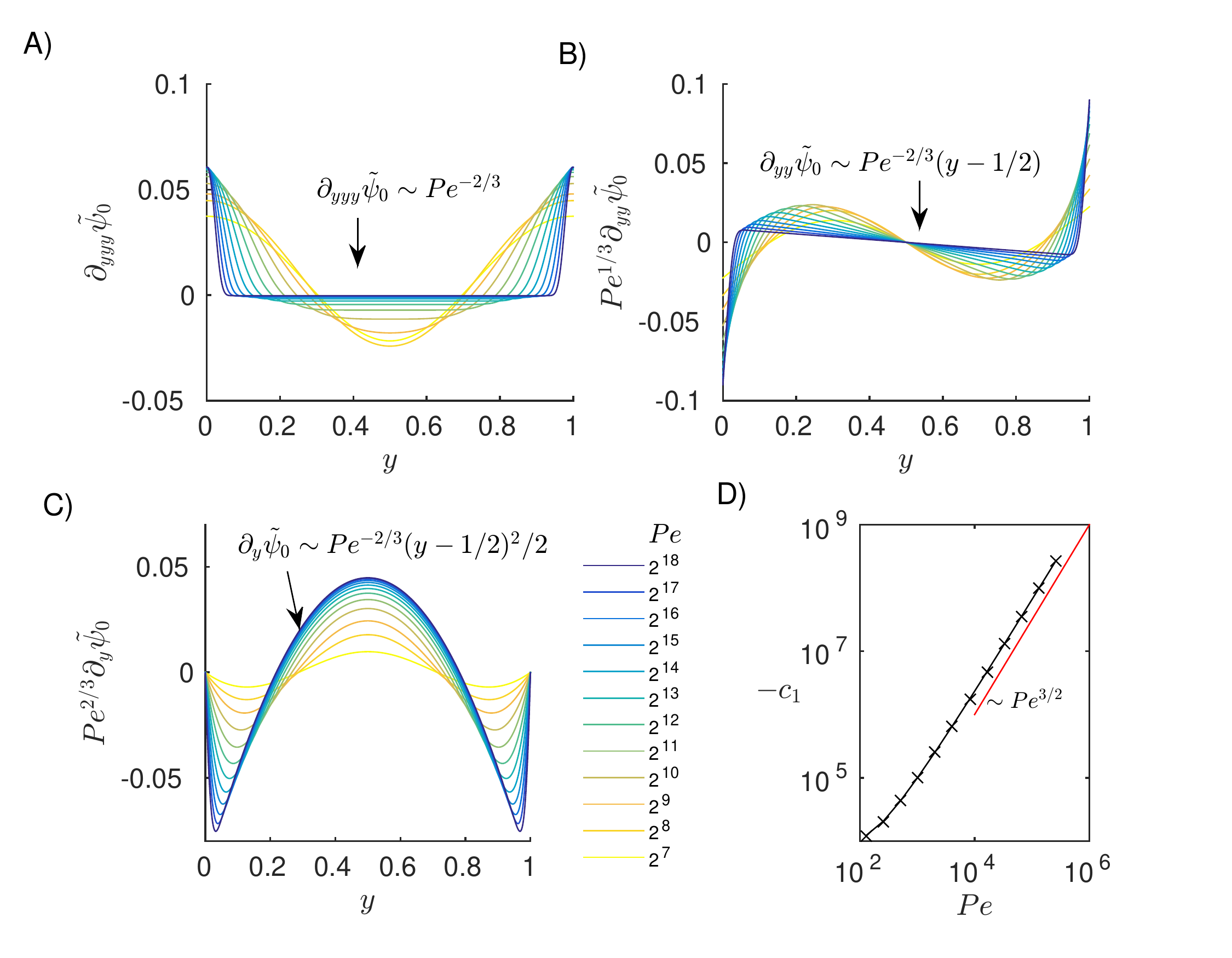}}
  \caption{Plots of derivatives of the approximate optimal flows $\tilde{\psi}_0 \equiv \tilde{\psi}_1/c_1$ along the channel
midline ($x = L/2$) scaled by powers
of $Pe$ to show convergence in the wall boundary layers near
$y = 0$ and 1. We have $\partial_y^3 \tilde{\psi}_0 \sim Pe^{0}$ (A),
$\partial_y^2 \tilde{\psi}_0 \sim Pe^{-1/3}$ (B), and
$\partial_y \tilde{\psi}_0 \sim Pe^{-2/3}$ (C). Outside
the boundary layers, the three functions scale like $Pe^{-2/3}$ times
polynomials of order 0 (A), 1 (B), and 2 (C). Each line
corresponds to a value of $Pe$ listed between panels C and D.
Panel D shows the scaling of the normalizing constant
defined in (\ref{ctilde}), $c_1 \sim Pe^{3/2}$.
}
\label{fig:PsiApproxFig}
\end{figure}

Therefore, at a given $x$ away from the upstream and downstream 
boundaries, we approximate (\ref{psi01tilde})--(\ref{psi02tilde}) by
\begin{align}
 &\mbox{PDE}  &   &\mbox{Top BCs}  &  &\mbox{Bottom BCs} \nonumber \\
&- 2\partial_y^4 \tilde{\psi}_0 = f,  & &\tilde{\psi}_0 = \partial_y \tilde{\psi}_0 = 0, & &\tilde{\psi}_0 = \partial_y \tilde{\psi}_0 = 0 \label{psi0atilde} 
\end{align}
\nn This is the equation for an elastic beam with clamped-clamped boundary conditions, loaded by sharp concentrations of force density ($-f/2$), equal
and opposite in the boundary layers at each end. In appendix
\ref{sec:elast} we find approximate scalings of $\tilde{\psi}_0$
and its derivatives inside and outside the boundary layers
by using a minimization principle from elasticity. Here we
give a brief explanation without the details. Since
$f$ has boundary layers of width
$\sim Pe^{-1/3}$ in $y$, we assume such boundary layers for
$\tilde{\psi}_0$ and its derivatives. In the boundary layer,
taking a $y$-derivative of $\tilde{\psi}_0$ is roughly equivalent
to multiplying by a factor of $Pe^{1/3}$. Since
$\partial_y^4 \tilde{\psi}_0 \sim f \sim Pe^{1/3}$ in
the boundary layer, we expect
$\partial_y^3 \tilde{\psi}_0 \sim Pe^{0}$, 
$\partial_y^2 \tilde{\psi}_0 \sim Pe^{-1/3}$,
$\partial_y \tilde{\psi}_0 \sim Pe^{-2/3}$ there.
These behaviors are shown in figure 
\ref{fig:PsiApproxFig}A-C. Outside the boundary layer,
we can find the scalings by the top and bottom
boundary conditions on $\partial_y \tilde{\psi}_0$ in
(\ref{psi0atilde}), which imply
\begin{align}
0 = \partial_{y} \tilde{\psi}_0|_{y = 1} - \partial_{y} \tilde{\psi}_0|_{y = 0} = 
\int_0^1 \partial_{yy} \tilde{\psi}_0 dy = 
\int_{\mbox{\small BL}} \partial_{yy} \tilde{\psi}_0 dy +
\int_{\mbox{\small outside BL}} \partial_{yy} \tilde{\psi}_0 dy. \label{scaling}
\end{align}
\nn The contribution inside the boundary layer
(denoted ``BL'' in (\ref{scaling})) has magnitude
$\sim Pe^{-1/3}$ in a region of width $\sim Pe^{-1/3}$.
To cancel this term, the contribution from outside
the boundary layer 
(denoted ``outside BL'' in (\ref{scaling})) is over a region
of width $\sim 1$, so $\partial_{yy} \tilde{\psi}_0$ should
have magnitude $\sim Pe^{-2/3}$ there. Since this region
has width $\sim 1$, it makes sense that the other 
$y$-derivatives of $\tilde{\psi}_0$ should also have
magnitude $\sim Pe^{-2/3}$ in the outer region, as
shown in figure 
\ref{fig:PsiApproxFig}A-C. Since $\partial^4_{y} \tilde{\psi}_0
\sim f \approx 0$ in this region, $\partial_y^3 \tilde{\psi}_0$, 
$\partial_y^2 \tilde{\psi}_0$, and
$\partial_y \tilde{\psi}_0$ are polynomials of degree 0, 1 and
2, respectively, also shown in figure 
\ref{fig:PsiApproxFig}A-C. In particular,
$\partial_{y} \tilde{\psi}_0$ is a parabola in this region, 
and with an appropriate prefactor $c_1$, 
$\tilde{\psi}_1 = c_1 \partial_{y} \tilde{\psi}_0$ can
cancel the parabolic flow of $\psi_{Pois}$ outside
the boundary layers, so the total approximate
optimal flow
$\partial_{y} \psi_1 = \psi_{Pois} + c_1 \partial_{y} \tilde{\psi}_0$ is
uniform in this region, with zero enstrophy expended.

Having discussed the scalings of $\tilde{\psi}_0$ and its
$y$-derivatives, we can compute $c_1$ from (\ref{ctilde})
written in terms of $\tilde{\psi}_0$, keeping only $y$-derivatives
\begin{align}
c_1 \approx  \pm2\sqrt{Pe^2 - 12L \psi_{top}^2} \Bigg/
\sqrt{\int_0^1 \int_0^L \left(\partial_y^2 \tilde{\psi}_0\right)^2 dx dy}. \label{ctilde1}
\end{align}
\nn In the denominator, the integral has a contribution
$\sim Pe^{-1}$ from the boundary layer and 
$\sim Pe^{-4/3}$ from outside, so it is
$\sim Pe^{-1}$. To determine the scaling of 
$c_1$ for the approximate optimal flow,
we need to know $\psi_{top}$ for this flow, defined
as $\psi_{top,opt}$. We know
$\psi_{top,opt}$ is bounded by $\bar{\psi}_{top}$, and
so the numerator of (\ref{ctilde1}) is $O(Pe)$ (unless
$\psi_{top,opt} \to \bar{\psi}_{top}$ as $Pe \to \infty$,
which does not agree with the numerical solutions).
Therefore we have $c_1 \sim Pe^{3/2}$, shown in figure 
\ref{fig:PsiApproxFig}D. This constant converts the scalings
for $\tilde{\psi}_0$ into those for the approximate optimal flow component $\tilde{\psi}_1$. For example,
$\partial_y \tilde{\psi}_1 = c_1 \partial_y \tilde{\psi}_0 \sim 
Pe^{3/2} Pe^{-2/3} = Pe^{5/6}$. The total approximate optimal
flow is $\psi_1 = \psi_{Pois} + \tilde{\psi}_1$, which requires
knowing the constant $\psi_{top}$ where the optimal heat transfer
occurs. We use the intuition from figure \ref{fig:FlowVsaFig}
that the optimal $\psi_{top}$ is that for which the core flow is
a uniform flow with zero enstrophy expended. By figure
 \ref{fig:PsiApproxFig}C, $\partial_y \tilde{\psi}_1$ has a
backward parabolic flow $\sim Pe^{5/6}$ outside the boundary
layers, so $\psi_{top}\sim Pe^{5/6}$ gives a parabolic flow
$\partial_y \psi_{Pois}$ which cancels this core flow up 
to a constant, giving a uniform core flow.

\begin{figure}
  \centerline{\includegraphics[width=17cm]
  {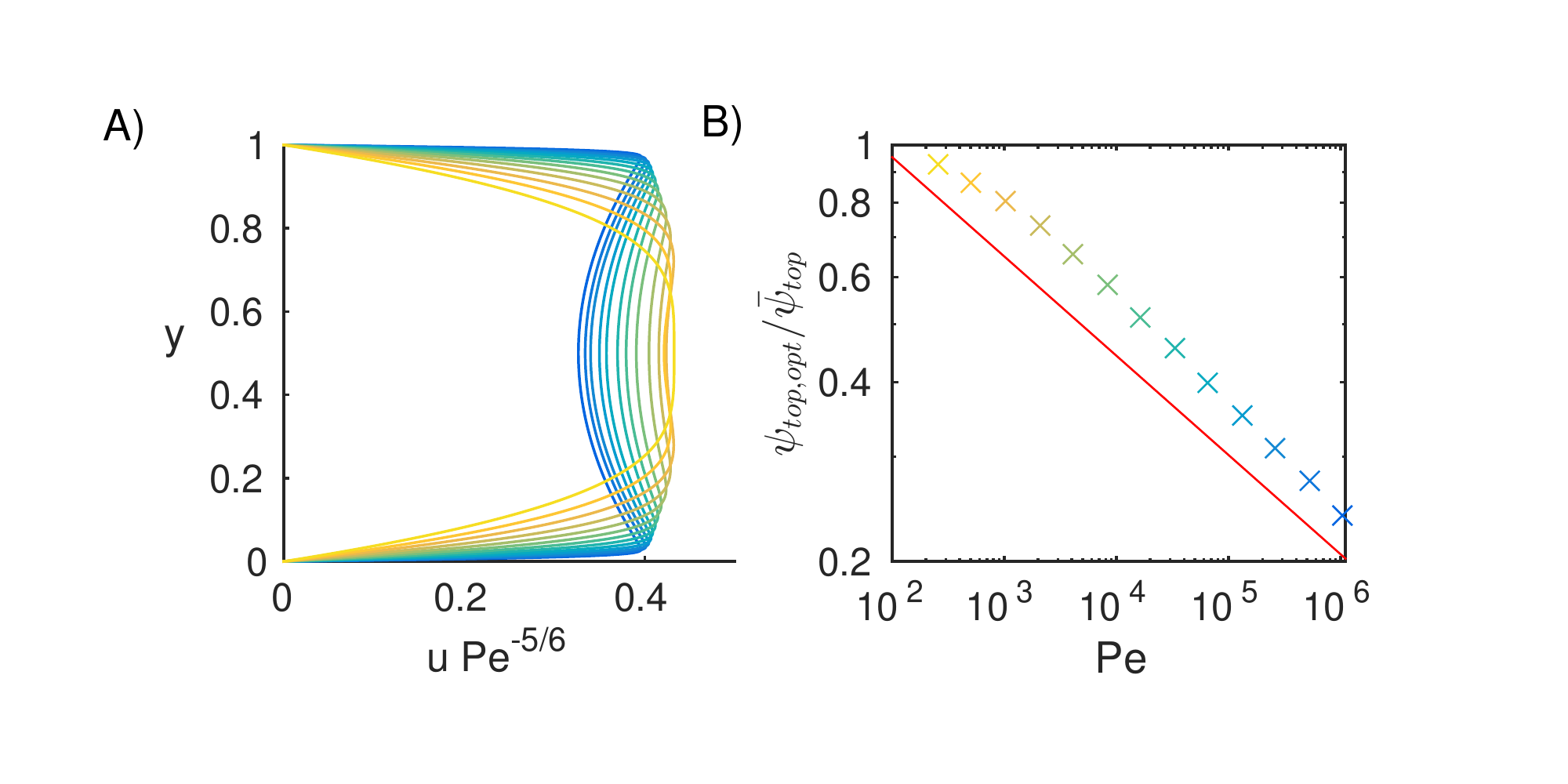}}
  \caption{Approximate optimal flows (A) at the 
$x$ midpoint of the channel for various $Pe = 2^8$ -- $2^{20}$,
rescaled by $Pe^{5/6}$. The colors correspond to $Pe$ values
marked by crosses in panel B, which also show how the optimal
flow rate $\psi_{top,opt}$ decreases relative to that of the starting Poiseuille flow solution, $\bar{\psi}_{top}$. The
red line gives the scaling $Pe^{-1/6}$. The channel
length is $L = 4$.}
\label{fig:PsiOptFig}
\end{figure}

In figure \ref{fig:PsiOptFig}A we plot the approximate optimal
flows $u = \partial_y \psi_1$ over $Pe = 2^8$ -- $2^{20}$ and find a good collapse when
divided by $Pe^{5/6}$. The corresponding values of $\psi_{top}/\bar{\psi}_{top}$, denoted $\psi_{top,opt}/\bar{\psi}_{top}$ are plotted in panel B. The red line shows
the expected scaling $Pe^{-1/6}$. In fact $\psi_{top,opt}/\bar{\psi}_{top}$ decreases
slightly faster. This is
reflected in panel A by the fact that the rescaled profiles show
a slight decrease with increasing $Pe$. Instead of uniform flow
in the core, there is a moderately concave parabolic flow there.

To compute the heat transferred
by these flows we need to integrate $\partial_y T_1$ over the walls, 
where $T_1$ solves (\ref{T1}).

\begin{figure}
  \centerline{\includegraphics[width=14cm]
  {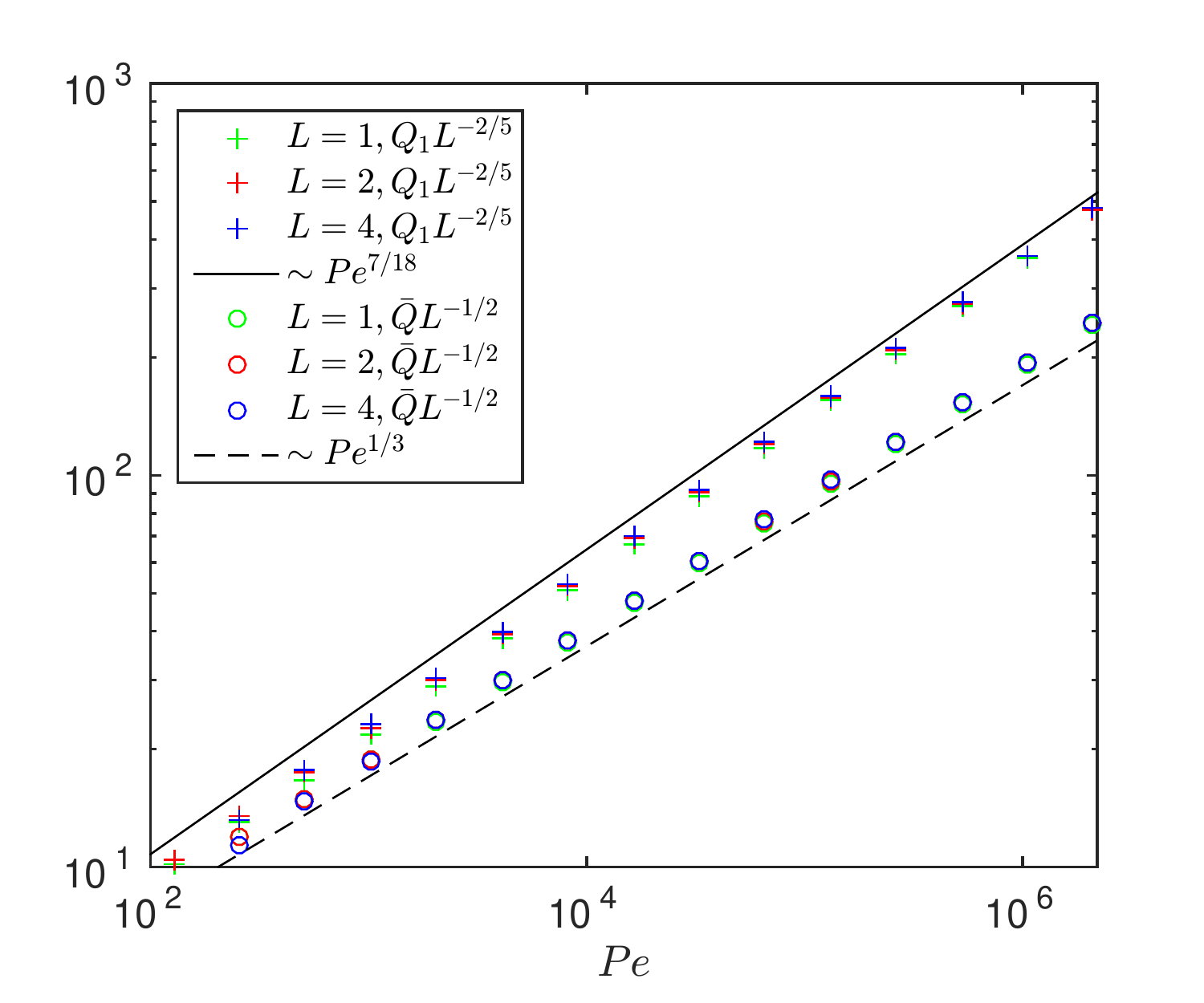}}
  \caption{Comparison of the heat transferred by the approximate optimal flows ($Q_1$, plusses) with
that transferred by the Poiseuille flow at the same value of $Pe$ ($\bar{Q}$, circles), at
three channel lengths ($L = 1, 2$, and 4) at 
$Pe$ ranging from $2^7$ -- $2^{21}$. The values are scaled
by factors of $L$ as described in the text.}
\label{fig:Q1Fig}
\end{figure}

\nn As mentioned earlier, the similarity solution for the temperature field in a Poiseuille flow has heat transfer given 
by \cite{shah2014laminar}:
\begin{align}
\bar{Q} \sim \int_0^L 1.23 \left(\frac{2 Pe}{x \sqrt{3L}}\right)^{1/3} dx
= 1.85 (4/3)^{1/6} Pe^{1/3} L^{1/2} \label{barQ}
\end{align}
\nn The quantity in parentheses in (\ref{barQ}) is proportional to 
the similarity 
variable $\eta$ raised to the -1/3 power, and is $x$ scaled by the
velocity gradient at the wall (i.e. the vorticity) for Poiseuille
flow, $\sim Pe/\sqrt{L}$. For the approximate optimal flow, the wall vorticity is proportional to $Pe^{7/6}$.
To find the scalings we substitute $Pe^{7/6}$ for
 $Pe/\sqrt{L}$ in (\ref{barQ}), and drop the constants and $L$ dependence
(which are altered for the approximate optimal flows). We obtain
$Q_1 \sim Pe^{7/18}$. The discussion of the $L$ dependence is deferred to section \ref{sec:L}. In figure \ref{fig:Q1Fig} we compare $Q_1$ with $\bar{Q}$,
the heat transferred by Poiseuille flow, at 
$Pe$ = $2^7$ -- $2^{21}$ and three channel lengths ($L = 1, 2,$ and 4). The
data approximately collapse when scaled by $L^{1/2}$ for Poiseuille flow (already discussed) and
$L^{2/5}$ for the approximate optimal flows (discussed later, in section \ref{sec:L}).
The improvement over Poiseuille flow reaches a factor of 2 near $Pe = 10^6$ for $L = 1$.

\section{Fully coupled system\label{sec:Full}}

The method of successive approximations in the last section has shown how
the initial temperature boundary layer $\sim Pe^{-1/3}$ produces the approximate
optimal flow boundary layer, and how that induces a sharper 
temperature boundary layer $\sim Pe^{-7/18}$ at the next iteration.
With this temperature boundary layer, the optimal flow is then modified
through the source term in (\ref{psi1}) at the next iteration. Proceeding in this way, one would hope
to determine the flow and temperature boundary layers 
that simultaneously solve the fully coupled problem (\ref{T})--(\ref{mu}).
Computationally it is difficult to achieve convergence with Newton's method
for  $Pe \gtrsim 10^5 - 10^6$, so
we proceed with a unidirectional flow approximation. We might expect
the optimal
unidirectional flow to be better than the optimal 2D flow field solution
in (\ref{T})--(\ref{mu}) because that flow is constrained to be Poiseuille flow at the upstream and downstream boundaries,
while the optimal unidirectional flow can take on a wider range of values at those locations. The optimal
2D flow field is nearly unidirectional in the middle of the channel, so assuming a unidirectional flow
may not sacrifice much. 

Let us approximate the optimal flow velocity as unidirectional,
$u(y) \mathbf{\hat{e}}_x$, with a boundary layer of thickness $\sim Pe^{-\alpha}$.
The flow $u(y)$ rises from 0 to a value $\sim Pe^{\beta}$ outside the
boundary layer, where the flow is assumed uniform, so the vorticity
is zero there. In the boundary layer, the vorticity
$\omega = \partial_y u \sim Pe^{\beta + \alpha}$.

One equation relating $\alpha$ and $\beta$ is that the total
enstrophy is $Pe^2$:
\begin{align}
Pe^2 = L \int_0^1 \omega^2 dy \sim Pe^{2\beta + \alpha}
\rightarrow 2\beta + \alpha = 2. \label{cond1}
\end{align}
\nn The thinner the boundary layer (larger $\alpha$), the smaller the velocity inside and outside the boundary layer (smaller $\beta$). 
Another equation comes from the condition that the temperature boundary
layer thickness should have the same scaling as the flow boundary layer
thickness. 
This is essentially what is expressed by equation (\ref{psi1}),
where the temperature boundary layer appears in 
$\nabla^\perp T \cdot \nabla m$ 
and creates a flow boundary layer in $\psi$. The relationship between
the boundary layers is also shown in figures \ref{fig:fApproxFig} and
\ref{fig:PsiApproxFig}. The condition is
optimal physically because it gives the fastest possible flow in
the temperature boundary layer, so it convects the most heat out of the
channel.
If instead the flow boundary layer were smaller than the temperature boundary layer (i.e. increased $\alpha$), 
by (\ref{cond1}) the flow speed magnitude
would be reduced everywhere (decreased $\beta$) including
the temperature boundary layer, 
so less heat would be convected out of the channel. If on the other
hand
the flow boundary layer were larger than the temperature boundary layer,
then enstrophy is being spent to create a faster core flow at the
expense of enstrophy in the boundary layer. Since
the temperature is uniformly zero in the core, there is no advantage to
having a faster flow there, while the slower boundary layer flow
convects less heat. The similarity solution tells us how the
temperature boundary layer is related to the flow boundary layer. In
the flow boundary layer, $u \sim Pe^{\beta + \alpha} y$. Assuming
this profile holds throughout the temperature boundary layer, we
can apply the similarity solution for the temperature field
in a linear velocity profile \cite{leveque1928laws}, giving
the temperature boundary layer thickness $\sim |$wall vorticity$|^{-1/3} = |\partial_y u|^{-1/3}$ $\sim  Pe^{-(\beta + \alpha)/3}$.
Setting this equal to the flow boundary layer thickness gives
\begin{align}
Pe^{-(\beta + \alpha)/3} \sim Pe^{-\alpha}
\rightarrow \beta = 2\alpha. \label{cond2}
\end{align}
\nn Combining (\ref{cond1}) and (\ref{cond2}) yields $\alpha = 2/5$ and
$\beta = 4/5$. 

\begin{figure}
  \centerline{\includegraphics[width=17cm]
  {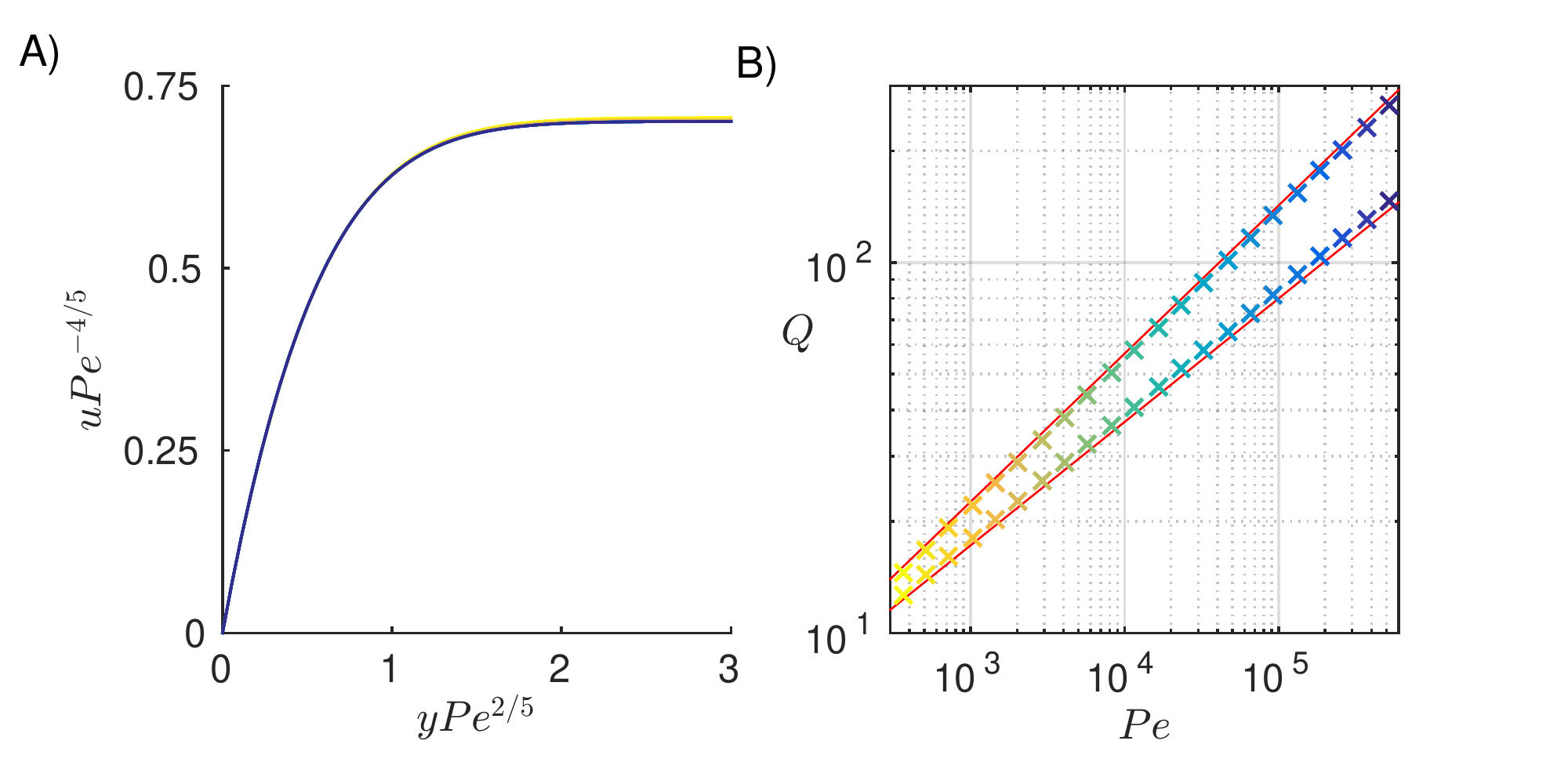}}
  \caption{Optimal unidirectional flows $u(y)$ and heat transferred. A)
For $Pe$ ranging from $2^{8.5}$ to $2^{19}$ in multiplicative increments of
$2^{0.5}$, the optimal unidirectional flow
profiles near the lower wall, with $u$ rescaled by $Pe^{4/5}$ and $y$ scaled by 
$Pe^{-2/5}$. B) The heat transferred by the optimal
unidirectional flow (upper set of crosses), with scaling line
$Pe^{2/5}$ (upper red line), and the heat transferred by the Poiseuille
flow at the same $Pe$ (lower set of crosses), with scaling
line $Pe^{1/3}$ (lower red line). Here $L$ = 1 and $\delta = 0.1$. 
The colors for the 22 lines in panel A correspond to the colors of
the crosses in panel B at the same $Pe$.}
\label{fig:EnstrophyPe}
\end{figure}

These values can also be derived another way. Instead of assuming
flow boundary layer thickness $\sim$ temperature boundary layer thickness, we
can allow it to arise naturally in the optimization. We want to maximize
$\partial_y T$ (actually its integral over $x$), 
and $T$ falls from 1 to 0 over the temperature
boundary layer, so $\partial_y T \sim$ (temperature boundary layer thickness)$^{-1}$. If the flow boundary layer is much larger than the temperature boundary layer, the temperature evolves in a linear shear flow,
and the boundary layer thickness
$\sim \omega_{wall}^{-1/3} \sim Pe^{-(\beta + \alpha)/3}\sim 
Pe^{-(1 + \alpha/2)/3}$ using (\ref{cond1}) to eliminate $\beta$.
If the flow boundary layer is much smaller 
than the temperature boundary layer, the temperature evolves in an
essentially 
uniform flow, with boundary layer thickness
$\sim$ (flow speed)$^{-1/2}$. This follows from the similarity solution for the temperature in a uniform flow field \cite{haase2015graetz, graetz1882ueber}, the same as the similarity solution to the heat equation in
one space and one time variable.  
We have boundary layer thickness $\sim$ (flow speed)$^{-1/2} \sim Pe^{-\beta/2} \sim 
Pe^{-(1 - \alpha/2)/2}$ using (\ref{cond1}) again to eliminate $\beta$.
At $\alpha = 2/5$, both expressions give the temperature boundary
layer thickness $\sim Pe^{-2/5}$, the same as the flow boundary layer. If
the flow boundary layer were smaller, then
$\alpha > (1 - \alpha/2)/2 \rightarrow (1 - \alpha/2)/2 \leq 2/5 $, so the
temperature boundary layer would be larger than $Pe^{-2/5}$. If the flow boundary layer
were larger, then $\alpha \leq (1 + \alpha/2)/3 \rightarrow (1 + \alpha/2)/3 \leq 2/5$, and again the
temperature boundary layer would be larger than $Pe^{-2/5}$. The optimum occurs when the
temperature boundary layer has the smallest thickness, $\sim Pe^{-2/5}$,
which is when the flow and temperature boundary layer thicknesses are equal.
In this case, the heat transfer scales as $Pe^{2/5}$, which is slightly
greater than that for the first iteration of the method of successive approximations ($Pe^{7/18}$), about 16\% greater at $Pe = 10^6$.
This slight difference may explain the close agreement between the exact and approximate heat transfer in figure \ref{fig:ApproxNewtonFig}.
The improvement of $Pe^{2/5}$ over the $Pe^{1/3}$ scaling for Poiseuille flow 
is a factor of about 2.5 at $Pe = 10^6$ with $L = 1$.

To test these hypotheses we employ a quasi-Newton (Broyden) method to solve
for the optimal $u(y)$ using equations (\ref{T})--(\ref{mu})
specialized to the case $\psi(x,y) \equiv \psi(y)$, with no boundary conditions
needed for $\psi$ at the $x$-boundaries. $T$ and $m$ are still functions of $x$ and $y$ with
the same boundary conditions.
Unlike Newton's method, Broyden's method uses a dense approximation to the Jacobian matrix, but it is manageable since $\psi(y)$ is
discretized on a 1D grid in $y$. An important savings comes from the fact that
$\partial_y^3 \psi(y)$ is constant outside the boundary layer, so we only need to explicitly compute
$\partial_y\psi(y)$ in a region slightly larger than the boundary layer, and can represent
$\partial_y\psi(y)$ outside by a quadratic polynomial that matches the solution
slightly outside the boundary layer. Only 90 unknowns are used in this version of
Broyden's method, though ill-conditioning is still an issue as the boundary layer
shrinks.

In figure \ref{fig:EnstrophyPe} we plot the results of the unidirectional flow
optimization. Panel A shows the flow profiles near the boundary layer for
$Pe$ ranging from $2^{8.5}$ to $2^{19}$ in multiplicative increments of
$2^{0.5}$, with
$u$ scaled by $Pe^{4/5}$ and $y$ scaled by 
$Pe^{-2/5}$, showing the expected behaviors. Panel B shows the heat transferred
by these flows (upper set of crosses), with scaling $Q \sim Pe^{2/5}$, compared
to the Poiseuille flow at the same $Pe$ (lower set of crosses), with scaling $Q \sim Pe^{1/3}$.

\section{Effect of varying $L$ \label{sec:L}}

We now discuss the dependence of solutions on the other
main parameter, $L$, the channel length/height. With fixed enstrophy,
as $L$ increases, the enstrophy spreads out, and therefore the flow
is weaker, which leads to a thicker temperature boundary layer. The
boundary layer is also thicker on average in the channel because it
spreads as $x$ increases (for $T$) or $L-x$ increases (for $m$). 
Therefore,
we expect the optimal solutions to have a thicker boundary layer and
slower flows at larger $L$, opposite to the effects of increasing $Pe$
with fixed $L$.

To determine how the optimal solutions scale with $L$ at fixed $Pe$, we can use two methods.
The first is the same as for the $Pe$ scalings: find power law exponents for
1. the boundary layer thickness and 2. the maximum flow speed such that 
1. the flow and temperature boundary layers have the same scaling law
and 2. total enstrophy scales as $Pe^2$,
i.e. is $L$-independent. Let us assume the the optimal flow and temperature fields
have boundary layer thicknesses $\sim L^{-\alpha}$, and that the
optimal flow speed has a maximum value $\sim L^{\beta}$.

With fixed enstrophy (fixed $Pe$),
\begin{align}
L^0 \sim L \int_0^1 \omega^2 dy \sim L^1L^{2(\beta + \alpha)}L^{-\alpha} \rightarrow 1 + \alpha +2\beta = 0.
\label{EnstrophycondL1}
\end{align}
\nn The temperature boundary layer thickness, which has the same
thickness as the flow
boundary layer via (\ref{psi1})), scales as 
(wall vorticity)$^{-1/3} \sim L^{(-\alpha-\beta)/3}$, and
spreads like $x^{1/3} \sim L^{1/3}$ in a channel of length $L$. Matching
this to the flow boundary layer thickness, 
\begin{align}
L^{(-\alpha-\beta)/3}L^{1/3} \sim L^{-\alpha} \rightarrow -\alpha/3-\beta/3 + 1/3 = -\alpha.
\label{EnstrophycondL2}
\end{align}
\nn Solving
(\ref{EnstrophycondL1}) and (\ref{EnstrophycondL2}) we have $\alpha = -3/5$ and
$\beta = -1/5$. 

\begin{figure}[h]
  \centerline{\includegraphics[width=17cm]
  {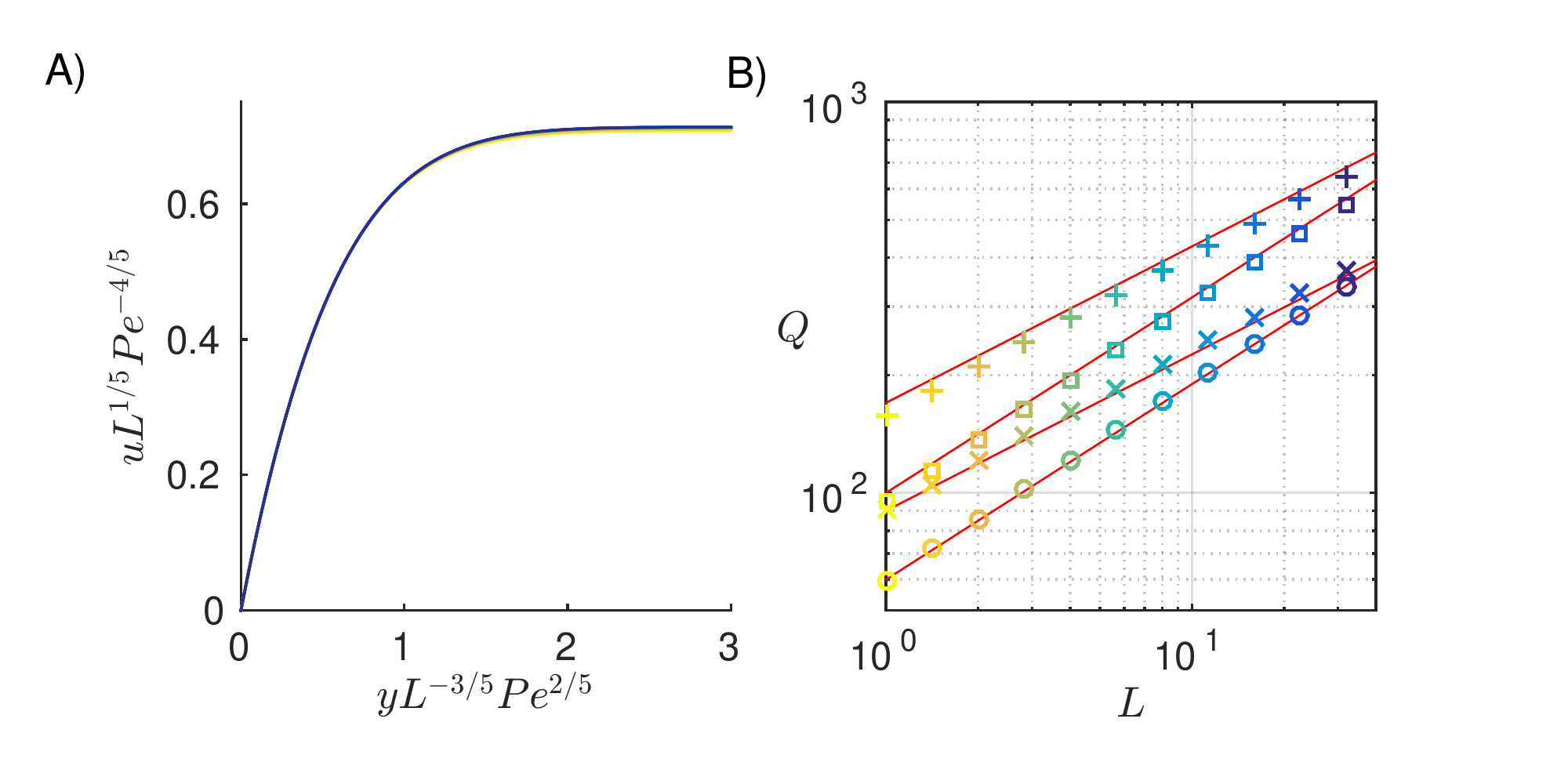}}
  \caption{Scalings of optimal unidirectional flows with $L$. A, Optimal unidirectional flows with fixed enstrophy at $L$ varying from 1 to 32 in factors of $\sqrt{2}$, at $Pe = 2^{15}$ and
$2^{17}$. B, Corresponding heat transferred $Q$ versus $L$ 
at $Pe = 2^{15}$ (crosses for
optimal flows and circles for Poiseuille flows) and $Pe = 2^{17}$ (plusses for
optimal flows and squares for Poiseuille flows). The red lines near the
optimal-flow data give the scaling $L^{2/5}$ while those near the
Poiseuille-flow data give the scaling $L^{1/2}$.}
\label{fig:EnstrophyLPe}
\end{figure}

Figure
\ref{fig:EnstrophyLPe}A shows the numerical solutions collapsed 
with the appropriate scalings in $L$ and $Pe$ applied
simultaneously. 
We vary $L$ from 1 to 32 in
factors of $\sqrt{2}$, and at two different $Pe$, $2^{15}$ and $2^{17}$,
yielding 22 curves which are collapsed in panel A.
Larger 
$L$ is challenging for computations because of the large grid needed
to cover the domain at a given resolution.
The scaling for $Q$ is given by
\begin{align}
Q = 2\int_0^L \partial_y T |_{y = 0} dx \sim \int_0^L \left(\frac{L^{(\alpha+\beta)/3}}{x}\right)^{1/3} dx \sim L^{2/5}. \label{QEnstrophyL}
\end{align}
\nn In (\ref{QEnstrophyL}) we have used the fact that the temperature boundary
layer thickness scales as (wall vorticity)$^{-1/3} \sim L^{(\alpha+\beta)/3}$,
and grows as $x^{1/3}$ in the linear-flow similarity solution.
In (\ref{QEnstrophyL}), the heat flux is proportional to the inverse of
the temperature boundary
layer thickness, so we are integrating the reciprocal of these terms.
The optimal heat flux scaling $Q \sim L^{2/5}$ is shown by the plusses and crosses in
figure \ref{fig:EnstrophyLPe}B, and compared with the $L^{1/2}$ scalings for
the starting uniform flows at the same $Pe$ and $L$ (squares and circles).
The two diverge as $L$ becomes smaller, where the boundary layers are thinner and
there is a greater advantage to concentrating the flow in the boundary layer.

There is a second way to determine the optimal scalings with $L$. It is
to recognize that the boundary
layers have no dependence on the channel height, $H$ (dimensional) or 1 
(dimensionless). Therefore, in the boundary layer regime, there is
only one relevant length scale in the problem, $L_0$ (dimensional channel length). By using $L_0$ instead of 
$H$ as the length scale, we can eliminate one dimensionless parameter,
$L$ say. So long as the boundary layer is much smaller than
$H$, the solutions do not depend on $L = L_0/H$, only
the energy/enstrophy budget, which needs to be redefined as $\tilde{Pe}$ instead of
$Pe$ when $L_0$ is the length scale. All the information is contained in the problem
as a function of $\tilde{Pe}$, so it should yield the scalings
with respect to $L$ and $Pe$ separately. Indeed, we can obtain the scalings using
this approach, as shown in Appendix \ref{Nondim}.


\section{Physical considerations \label{sec:Physical}}

We have given results in terms of 
$Pe$, similarly to 
\cite{hassanzadeh2014wall,souza2015transport,goluskin2016bounds,tobasco2016optimal,alben_2017}, because this measures the energy of the flow,
the key constraint in our problem. To estimate
the physical scales of these flows, 
we now list the Reynolds numbers corresponding to the flows we have
computed for the most common convecting fluid, air.

For the case of fixed enstrophy = $Pe^2$, the dimensionless Poiseuille
flow $u(y) = 6 Pe (y - y^2)/\sqrt{12 L}$ has $y$-averaged flow speed
$\langle u \rangle = Pe/\sqrt{12 L}$. The corresponding dimensional
flow speed is  $\langle u \rangle \kappa/H$. Using this
flow speed and the channel height $H$ as a typical length,
we define the Reynolds number as 
\begin{align}
Re \equiv \frac{Pe}{\sqrt{12 L}}\frac{\kappa}{H}\frac{H}{\nu} = \frac{Pe}{\sqrt{12 L}}
\frac{1}{Pr}.
\end{align}
\nn where $Pr = \nu/\kappa$ is the Prandtl number, about
0.8 for air at 300K. Therefore $Pe = 10^5$ corresponds to
$Re = 10^4-3 \times 10^4$ for Poiseuille flow with
$L$ ranging from 1 to 8. The optimal flows shown in 
figure \ref{fig:EnstrophyLPe}A grow more slowly with increasing $Pe$ and decrease more slowly
with increasing $L$: $\langle u \rangle \approx 0.7 L^{-1/5}Pe^{4/5}$.
At $Pe = 10^5$, $Re = 4000 - 7000$ for 
$L = 1 - 8$. At $Pe = 10^5$ we have significant improvement
in heat transfer, about a factor of $1.6$ over Poiseuille flow at $L = 1$.
The corresponding $Re$ are near 
the critical values for the transition to turbulence in
plane Poiseuille flow, $Re$ = 5772 for the linear instability, about 2900 for finite-amplitude
disturbances, and about 1000 in experiments due to 3D effects \cite{schmid2012stability}. These
values will undoubtedly change for the optimal flow profiles. 
Since the instability is convective, the transition may occur sufficiently slowly that the flow remains laminar
in the moderate-length channels considered here \cite{schmid2012stability,shah2014laminar} and in heat sink
geometries \cite{hamburgen1986optimal}. 

A steady unidirectional flow $u(y)$ is produced by a pressure gradient
$\partial_x p = \nu \partial_{yy} u$. For the optimal flows with fixed 
enstrophy in 
figure \ref{fig:EnstrophyLPe}A, this function has a maximum at the wall
and decreases to zero outside the boundary layer. One could 
achieve an approximation to this pressure gradient with
fans $\sim Pe^{-2/5}$ in size placed adjacent to the wall
at the upstream and downstream boundaries, to produce the boundary layers
of the appropriate sizes.

%
%
%
%
%


\section{Conclusion \label{sec:Conc}}

In this paper we have computed sequences of optimal flows for heat transfer in a channel.
They are only locally optimal, and proceed from a particular initial solution: Poiseuille
flow. However, they give a lower bound on the improvement 
that can be obtained over this fundamental flow. 

Poiseuille flow
is perhaps the most common laminar flow in a duct or channel, the result of a uniform 
pressure gradient. The sequence of steady 2D flows we have computed starting from Poiseuille flow
are optimal under the constraint of a fixed rate of viscous
dissipation, $Pe^2$, equal to the power needed to pump the fluid through the channel. They are well-approximated by the optimal unidirectional flows we computed,
which have a boundary
layer of thickness $\sim Pe^{-2/5}$ where the flow rises sharply to a maximum 
speed $\sim Pe^{4/5}$ where the boundary layer meets the core flow. In the core, the flow
is uniform, no energy is dissipated, and the fluid temperature is zero, so it does not
carry any heat out of the channel. These results can be obtained by the simple
condition that the flow and temperature field should have boundary layer thicknesses of the same order at the optimum.

Using a decoupled approximation, we have shown mathematically 
how the temperature boundary
layer produces an increased flow near the boundaries in the optimal solutions, by
depleting the parabolic flow from the initial Poiseuille flow in the core of the channel.

A good approximation to the optimal flows can be produced simply, by changing the
uniform pressure gradient of Poiseuille flow to one that is localized in the boundary layer,
with unidirectional forcing (e.g. fans) localized in the boundary layer. In an appendix we show how to compute the 2D forcing field that corresponds to any given flow field as the solution to coupled Poisson problems.

We have also shown that in a channel of aspect ratio (length/height) $L$,
the boundary layer thickness scales as $L^{3/5}$ and the outer flow speed scales
as $L^{-1/5}$ for the optimal unidirectional flow. The results are obtained by again
matching the temperature and flow boundary layer thicknesses for the optimal solution,
or by using dimensional analysis (in an appendix), noting that the boundary layer
solutions do not depend on the channel height.

We have found a 60\% improvement in heat transfer over Poiseuille flow at the same
pumping power, at Reynolds numbers where the Poiseuille flow becomes unstable, using
air as the convecting fluid. This value is for aspect ratio $L = 1$; the improvement increases with shorter channels. In future work these approaches can be extended 
to unsteady 3D flows, which show promise for further heat transfer improvements
\cite{rips2017efficient}.


\begin{acknowledgments}
We acknowledge helpful discussions with 
Ari Glezer, Rajat Mittal, and Charles Doering. 
\end{acknowledgments}

\appendix

\section{Solving for $\mathbf{f}$ and $p$ \label{compf}}

We can compute $\mathbf{f}$ and $p$ from (\ref{NS}) given $\mathbf{u}$. We use the
Helmholtz-Hodge decomposition to write
\begin{align}
\mathbf{f} = -\nabla \phi + \nabla^\perp q.
\end{align}
The irrotational part $-\nabla \phi$ can be absorbed into the pressure gradient
in (\ref{NS}). We are left with the incompressible part $\mathbf{f} = \nabla^\perp q$. 
Inserting into (\ref{NS}) we have, for steady flows
\begin{align}
\rho\mathbf{u}\cdot\nabla\mathbf{u} = -\nabla p + \mu \Delta \mathbf{u} + \nabla^\perp q. \label{NS1}
\end{align}
\nn Taking the divergence of (\ref{NS1}) and using $\nabla \cdot \mathbf{u} = 0$ 
results in a Poisson problem for $p$:
\begin{align}
-\Delta p = \rho  \nabla \cdot \left(\mathbf{u}\cdot\nabla\mathbf{u} \right), \label{peq}
\end{align}
\nn and taking $\nabla^\perp \cdot$ (\ref{NS1}) gives a Poisson problem for $q$:
\begin{align}
-\Delta q = -\mu \Delta \omega + \rho \mathbf{u}\cdot\nabla\omega.\label{qeq}
\end{align}
\nn where $\omega = \nabla^\perp \cdot \mathbf{u}$ is the vorticity.
When the velocity field is known, so are the right hand sides of (\ref{peq}) and (\ref{qeq}).
Coupled Neumann and Dirichlet boundary conditions for (\ref{peq}) and (\ref{qeq}) are obtained by taking 
the normal and tangential components of 
(\ref{NS1}) at the boundaries:
\begin{align}
\hat{\mathbf{n}} \cdot \left(\rho\mathbf{u}\cdot\nabla\mathbf{u} - \mu \Delta \mathbf{u} \right)
&= -\partial_n p + \partial_s q. \label{BC1} \\
\hat{\mathbf{s}} \cdot \left(\rho\mathbf{u}\cdot\nabla\mathbf{u} - \mu \Delta \mathbf{u} \right)
&= -\partial_s p - \partial_n q. \label{BC2} 
\end{align}
\nn where $s$ and $n$ denote coordinates tangential and normal to the boundary. 
This is analogous to the derivation of Neumann pressure boundary conditions in
\cite{gresho1987pressure}.

\section{Estimating the boundary layer scalings by
minimizing the elastic energy of a beam \label{sec:elast}}

The beam equation
arises from the principle of minimum total potential energy
in elasticity \cite{landau1986te,renton2002elastic}. At a given $x$ away from the $x$ boundaries, our flow is analogous to a beam with deflection
$\tilde{\psi}_0(y)$ with bending modulus 2 and force density $-f$, which
minimizes 
\begin{align}
U_{elastic} = \int_0^1 (\partial_y^2 \tilde{\psi}_0)^2 + f \tilde{\psi}_0 \, dy \label{elast}
\end{align}
\nn The first term in (\ref{elast}) is the internal elastic energy due
to beam curvature, and also our enstrophy constraint neglecting
$x$-dependence of $\tilde{\psi}_0$. The second term in (\ref{elast}) is minus the work done by $f$ (and corresponds to the advection-diffusion
equation constraint for our flow). Using the clamp boundary conditions
in (\ref{psi0atilde}), 
\begin{align}
U_{elastic} = \int_0^1 \left[(\partial_y^2 \tilde{\psi}_0)^2 + f \int_0^{y} \int_0^{y'} \partial_{y''}^2 \tilde{\psi}_0 dy'' dy' \right] \,dy. \label{elast1}
\end{align}
\nn We assume $\partial_y^2 \tilde{\psi}_0 \sim Pe^\alpha$ in the
boundary layer and find $\alpha$ by assuming that both terms 
in $U_{elastic}$ are of the same order when $U_{elastic}$ is minimized. Using
$f \sim Pe^{1/3}$ and that each integration over the boundary
layer gives a factor of $Pe^{-1/3}$, the two terms in
(\ref{elast1}) become
\begin{align}
Pe^{2\alpha}Pe^{-1/3} \sim Pe^{1/3} Pe^\alpha Pe^{-1} \label{elast2}
\end{align}
\nn which implies $\alpha = -1/3$. 
In figure \ref{fig:PsiApproxFig}a--c we plot
$\partial_y^3 \tilde{\psi}_0$, $\partial_y^2 \tilde{\psi}_0$, and
$\partial_y \tilde{\psi}_0$ respectively along the channel midline
($x = L/2$)
for $L = 4$ and various $Pe$ (listed between panels c and d).
We find that each function has a boundary layer behavior
consistent with what we've described. In panel
b, we find $\partial_y^2 \tilde{\psi}_0 \sim Pe^{-1/3}$ in
the boundary layer. Differentiating with respect to $y \sim
Pe^{-1/3}$ in the boundary layer 
yields $\partial_y^3 \tilde{\psi}_0 \sim Pe^{0}$
(panel a), and integrating yields 
$\partial_y \tilde{\psi}_0 \sim Pe^{-2/3}$
(panel c). These panels also show that outside the boundary
layers, all three functions scale like $Pe^{-2/3}$ times
polynomials of order 0 ($\partial_y^3 \tilde{\psi}_0$, panel a), 1
($\partial_y^2 \tilde{\psi}_0$, panel b), and 2
($\partial_y \tilde{\psi}_0$, panel c). The polynomial orders
are consistent with 
$\partial_y^4 \tilde{\psi}_0 \sim f \approx 0$ outside the
boundary layers.

To understand the scalings outside the boundary
layers we write an approximation for
$\partial_y^2 \tilde{\psi}_0$, valid inside and outside
the boundary layers, as $a_1(\eta) Pe^{-1/3} + a_2(y) Pe^{\beta}$.
Plugging this expression into $U$, we have an increased energy
outside the boundary layer with a decreased energy inside the
boundary layer, if the sign of $a_2(0)$ is opposite that of 
$a_1(\eta)$ as $\eta \to \infty$. Both energies are of the same order if
$\beta = -2/3$, and a net energy decrease occurs for an
appropriate choice of $a_2(y)$. 

\section{$L$ scalings using alternative nondimensionalization \label{Nondim}}

We define a new system of dimensionless quantities with $L_0$ as
the characteristic length instead of $H$, and denote them with tildes. 
First, we relate $\tilde{Pe}$ to $Pe$. With fixed enstrophy,
\begin{align}
Pe^2 = \frac{\dot{E} H^2}{W \mu \kappa^2} = \frac{\dot{E} L_0^2}{W \mu \kappa^2}
\frac{H^2}{L_0^2} = \tilde{Pe}^2 L^{-2}.
\end{align}
\nn The scalings for the optimal flow boundary layer thickness $\tilde{q}$,
maximum flow speed $\tilde{U}$, and heat transferred $\tilde{Q}$ with respect to $\tilde{Pe}$ are the same as the scalings of the corresponding quantities
($q,U,Q$) with respect to 
$Pe$ with any fixed value of $L$ (that yields boundary-layer
solutions):
\begin{align}
\frac{q}{L} &= \frac{q_d}{H} \frac{H}{L_0} = \frac{q_d}{L_0} = \tilde{q} \sim \tilde{Pe}^{-2/5} =  Pe^{-2/5}L^{-2/5} \rightarrow q \sim Pe^{-2/5}L^{3/5} \\
UL &= \frac{U_d H}{\kappa} \frac{L_0}{H} = \frac{U_d L_0}{\kappa} = \tilde{U} \sim \tilde{Pe}^{4/5} =  Pe^{4/5}L^{4/5} \rightarrow U \sim Pe^{4/5}L^{-1/5} \\
Q &= \tilde{Q} \sim \tilde{Pe}^{2/5} = Pe^{2/5}L^{2/5}.
\end{align}
\nn We have denoted the dimensional quantities by subscript $d$, which
allows us to translate between the dimensionless sets ($q,U,Q, Pe$) and
($\tilde{q},\tilde{U},\tilde{Q}, \tilde{Pe}$).


These are the same scalings as in Section \ref{sec:L}.

\bibliographystyle{unsrt}
\bibliography{OptimalChannelFlow}

\end{document}